\documentclass[aps,prc,twocolumn,floatfix,superscriptaddress,amsmath,amssymb,nofootinbib]{revtex4-2}

\usepackage{dcolumn}
\usepackage{float}
\usepackage{graphicx}
\usepackage[colorlinks=true,linkcolor=blue,citecolor=blue,urlcolor=blue]{hyperref}
\usepackage{orcidlink}
\usepackage{microtype}

\begin{document}

\title{Exactness of the normal-ordered two-body truncation of three-nucleon forces}

\author{Maxwell~Rothman\orcidlink{0000-0001-9991-5670}}
\affiliation{Department of Physics and Astronomy, University of
  Tennessee, Knoxville, Tennessee 37996, USA}

\author{Ben~{Johnson-Toth}\orcidlink{0009-0003-5525-0357}}
\affiliation{Department of Physics and Astronomy, University of
  Tennessee, Knoxville, Tennessee 37996, USA}  

\author{Francesca~Bonaiti\orcidlink{0000-0002-3926-1609}}
\affiliation{Facility for Rare Isotope Beams, Michigan State University, East Lansing, Michigan 48824, USA}
\affiliation{Physics Division, Oak Ridge National Laboratory, Oak
  Ridge, Tennessee 37831, USA}

\author{Gaute~Hagen\orcidlink{0000-0001-6019-1687}}
\affiliation{Physics Division, Oak Ridge National Laboratory, Oak
  Ridge, Tennessee 37831, USA}
\affiliation{Department of Physics and Astronomy, University of
  Tennessee, Knoxville, Tennessee 37996, USA}

\author{Matthias~Heinz\orcidlink{0000-0002-6363-0056}}
\affiliation{National Center for Computational Sciences, Oak Ridge National Laboratory, Oak Ridge, Tennessee 37831, USA}
\affiliation{Physics Division, Oak Ridge National Laboratory, Oak
  Ridge, Tennessee 37831, USA}

\author{Thomas~Papenbrock\orcidlink{0000-0001-8733-2849}}
\affiliation{Department of Physics and Astronomy, University of
  Tennessee, Knoxville, Tennessee 37996, USA}
\affiliation{Physics Division, Oak Ridge National Laboratory, Oak
  Ridge, Tennessee 37831, USA}

\begin{abstract}
Reference-state-based many-body methods  start from Hamiltonians that are normal ordered with respect to the reference state. In low-energy nuclear physics applications normal-ordered Hamiltonians consisting of two- and three-nucleon forces are usually truncated at the two-body rank with residual three-nucleon operators being discarded. Benchmark computations have shown that this truncation is accurate, but we lack an understanding about why it works. We show that the normal-ordered two-body truncation is exact for zero-range three-body forces when nuclei are computed using the coupled cluster with singles and doubles method. 
As the nuclear three-nucleon force is short ranged and a three-body contact is a leading term in effective field theories of quantum chromodynamics, our result provides an analytical basis for the popular normal-ordered two-body approximation. 
\end{abstract}
\maketitle

\emph{Introduction.}
In the past two decades \emph{ab initio} computations of atomic nuclei have advanced from light nuclei to $^{208}$Pb~\cite{hu2022,hebeler2023,miyagi2024}. These computations start from effective field theories of quantum chromodynamics and aspire to solve the nuclear many-body problem using only controlled approximations~\cite{hergert2020,Ekstrom:2022yea}. The corresponding Hamiltonians include three-nucleon forces: a three-body contact~\cite{bedaque1999} in the case of  pionless effective field theory~\cite{bedaque2002,platter2005,kirscher2010,barnea2015,kirscher2015,lensky2016,bansal2018,Konig:2019xxk} and additional one- and two-pion exchange forces~\cite{epelbaum06} in the case of chiral effective field theories~\cite{epelbaum2009,machleidt2011,ekstrom2015a,reinert2018,Piarulli:2019cqu}. 

For wave-function-based \emph{ab initio} methods~\cite{dickhoff2004,soma2008,barrett2013,hagen2014,hergert2016,stroberg2019} the inclusion of three-nucleon forces is particularly expensive because of the large number of matrix elements that need to be produced, stored, and processed. While clever ways to compute these matrix elements are available~\cite{hebeler2015b,hebeler2021,takayuki2022,tichai2024}, their inclusion in many-body computations remains expensive. 

This is why the normal-ordered two-body approximation~\cite{hagen2007a,roth2012,ripoche2020,hebeler2023} is so popular in many-body methods that start from a reference state. In those computations the first step is to normal order the Hamiltonian with respect to a nontrivial reference state. This could be a product state from a mean-field computation or a phenomenological product state resulting from filling harmonic oscillator orbitals according to the nuclear shell model~\cite{mayer1955}. In normal-ordered Hamiltonians, the three-nucleon potentials contribute to the energy expectation value of the reference state, to the normal-ordered one- and two-body potentials, and as the ``residual'' three-nucleon potentials. The normal-ordered two-body approximation discards the latter. 

This approximation has become standard in \emph{ab initio} computations of atomic nuclei. Examples include the coupled-cluster method~\cite{hagen2007a,hagen2014,roth2012,binder2013b}, self-consistent Green's function approaches~\cite{cipollone2013,carbone2013b}, the in-medium similarity renormalization group~\cite{hergert2016,stroberg2019,heinz2021}, and the quantum Monte Carlo method~\cite{arthuis2023}. A variant of this approximation, where the third nucleon in the three-body force is integrated over with the density of a free Fermi gas and renders it a density-dependent two-body force~\cite{holt2009} is also popular~\cite{hagen2012a,hagen2012b,carbone2013}. Perhaps surprisingly, we know only little about the quality of the normal-ordered two-body approximation. 
Numerical results for $^4$He~\cite{hagen2007a} and $^4$He, $^{16}$O, and $^{40}$Ca~\cite{roth2012} demonstrated that this approximation is accurate when using the coupled-cluster method. However, what is really lacking are analytical insights about why this approximation works and in what sense it is controlled. 

It is the purpose of this Letter to fill this gap. We will show that the  truncation of three-nucleon Hamiltonians at the normal-ordered two-body rank is exact for zero-range three-nucleon potentials when nuclei are computed using the coupled-cluster with singles and doubles (CCSD) method. As we will see, this is a rare case where exact results are obtained for the nuclear many-body system. In what follows we formulate the problem, demonstrate the correctness of our claim, and show computational results that illustrate the accuracy of the CCSD approximation.

\emph{Statement of the problem.}
Hamiltonians from effective field theories~\cite{bedaque2002,epelbaum2009,Hammer:2019poc} of quantum chromodynamics have the form
\begin{align}
    \hat{H} &= \sum_{pq} \tau_q^p\hat{a}_p^\dagger \hat{a}_q + {1\over 4}\sum_{pqrs}v^{pq}_{rs}\hat{a}_p^\dagger \hat{a}_q^\dagger \hat{a}_s \hat{a}_r \nonumber\\
    &+{1\over 36}\sum_{pqrstu}W^{pqr}_{stu}\hat{a}_p^\dagger \hat{a}_q^\dagger \hat{a}_r^\dagger \hat{a}_u\hat{a}_t\hat{a}_s \ .
\end{align}
Here, $\tau_q^p$ denote the matrix elements of the kinetic energy, $v^{pq}_{rs}$ are two-body matrix elements, and $W^{pqr}_{stu}$ are three-body matrix elements. We use the convention that lower and upper indices correspond to incoming and outgoing states, respectively. The operators $\hat{a}_p^\dagger$ and $\hat{a}_p$ create and annihilate a nucleon in the single-particle state  $|p\rangle=\hat{a}_p^\dagger|0\rangle$, respectively. They fulfill the usual anticommutation relations for fermions. 

The reference state is the product state
\begin{equation}
\label{ref}
    |\phi\rangle = \prod_{i=1}^A\hat{a}^\dagger_i|0\rangle \ .
\end{equation}
Here and in what follows we use the convention that indices $i,j,k,\ldots$ refer to single-particle states occupied in the reference state while $a,b,c,\ldots$ refer to unoccupied single-particle states; indices $p,q,r,\ldots$ refer to any single-particle state. After normal ordering the Hamiltonian becomes
\begin{align}
\label{HamNO}
    \hat{H} &= E_0 + \sum_{pq} F_q^p\left\{\hat{a}_p^\dagger \hat{a}_q \right\}+ {1\over 4}\sum_{pqrs}V^{pq}_{rs}\left\{\hat{a}_p^\dagger \hat{a}_q^\dagger \hat{a}_s \hat{a}_r\right\} \nonumber\\
    &+{1\over 36}\sum_{pqrstu}W^{pqr}_{stu}\left\{\hat{a}_p^\dagger \hat{a}_q^\dagger \hat{a}_r^\dagger \hat{a}_u\hat{a}_t\hat{a}_s\right\} \ .
\end{align}
Here, the brackets $\{\cdot\}$ indicate normal ordering with respect to the reference state, i.e., operators that annihilate the reference state~(\ref{ref}) are to the right of those that do not. 
In Eq.~(\ref{HamNO})
\begin{align}
\label{NO-matele}
E_0 &=\sum_i \tau_i^i + {1\over 2}\sum_{ij}v_{ij}^{ij} + {1\over 6}\sum_{ijk}W_{ijk}^{ijk} \ , \nonumber\\ 
F_q^p &= \tau_q^p + \sum_i v_{iq}^{ip} +{1\over 2} \sum_{ij}W^{ijp}_{ijq} \ , \nonumber\\ 
V^{pq}_{rs} &= v^{pq}_{rs} + \sum_i W_{irs}^{ipq}
\end{align}
are the energy expectation value of the reference state, the Fock matrix, and the normal-ordered two-body matrix elements, respectively. We see that three-body matrix elements enter the zero-, one-, and two-body matrix elements~(\ref{NO-matele}). In the normal-ordered two-body approximation, the Hamiltonian~(\ref{HamNO}) is truncated at rank two; i.e., the last line in Eq.~(\ref{HamNO}) is discarded. One can easily generalize the normal-ordered two-body approximation to situations where a more complicated reference state enters by replacing the sums over hole states in Eq.~(\ref{NO-matele}) by sums over all single-particle states weighted by occupation numbers (see, e.g., Ref.~\cite{hergert2016}). 

We claim that the normal-ordered two-body truncation becomes exact for zero-range three-nucleon forces  when nuclei are computed with the coupled-cluster singles and doubles (CCSD) method.

\emph{Proof.}
To assert this claim we use a single-particle basis that comprises a lattice in position space. Such a basis has been used in nuclear lattice effective field theory~\cite{lee2009,lahde2019}. The single-particle states can be written as $|p\rangle \equiv |x_p, y_p, z_p, s_p, \tau_p\rangle$ where $x_p, y_p, z_p\in \{1, 2,\dots,L\}$ are integers denoting the lattice site, $s_p=\pm 1/2$ denotes the nucleon's spin projection, and $\tau_p=\pm 1/2$ denotes its isospin projection. Thus, the single-particle basis consists of $4L^3$ states. 

We start by considering even-even $N=Z$ nuclei. For light nuclei one usually thinks that such nuclei consist of  $\alpha$-particle clusters. Let us consider $^4$He, the simplest such nucleus. The reference state consists of four nucleons occupying the same lattice site. We note that this reference state certainly has nonzero overlap with the $^4$He ground state and is therefore suitable. 

It is now important to realize that only matrix elements of the form $W_{abc}^{def}$ and $W_{ijk}^{lmn}$ do not vanish [recall our convention stated in the text below Eq.~(\ref{ref})]. This is because the three-nucleon force is limited to nucleons residing on a single lattice site and the hole space consists of all states on a single lattice site (and the particle space consists of the complement). Thus, there is no lattice site that contains both particle and hole states. In particular the matrix elements $W_{abc}^{ijk}$ that directly contribute  to the correlation energy (see Eq.~(12) in Ref.~\cite{hagen2007a}) also vanish. 
The matrix elements $W_{ijk}^{lmn}$ enter the normal-ordered zero-, one-, and two-body matrix elements~(\ref{NO-matele}) and thereby do contribute to the ground-state energy. However, they do not contribute to the equations that determine the cluster amplitudes of the CCSD method.  To see this, one might either take a look at the corresponding equations in Ref.~\cite{hagen2007a} (which is admittedly somewhat tedious) or follow the arguments presented below.

Coupled-cluster theory is based on the similarity transformed Hamiltonian
\begin{equation}
    \overline{H}\equiv e^{-\hat{T}}\hat{H} e^{\hat{T}} \ , 
\end{equation}
where the cluster operator
\begin{align}
\label{cluster}
    \hat{T} = \sum_{ia}t_i^a\hat{a}^\dagger_a\hat{a}_i + 
    {1\over 4} \sum_{ijab}t_{ij}^{ab}\hat{a}^\dagger_a\hat{a}^\dagger_b\hat{a}_j\hat{a}_i +\ldots
\end{align}
consists of one-particle--one-hole, two-particle--two-hole excitations, and so on. For an $A$-body system the expansion~(\ref{cluster}) terminates at $A$-particle--$A$-hole excitations, but the commonly used CCSD approximation truncates after the first two terms. The unknown amplitudes $t_i^a$ and $t_{ij}^{ab}$ result from the solution of the following set of equations:
\begin{align}
\label{ccsd}
    \overline{H}_i^a&\equiv \langle\phi|\hat{a}^\dagger_i\hat{a}_a \overline{H}|\phi\rangle = 0 \ , \nonumber\\
    \overline{H}_{ij}^{ab}&\equiv \langle\phi|\hat{a}^\dagger_i\hat{a}^\dagger_j\hat{a}_b\hat{a}_a \overline{H}|\phi\rangle = 0 \ .
\end{align}
The matrix elements~(\ref{ccsd}) can be computed via the Baker-Campbell-Hausdorff expansion
\begin{align}
    \overline{H} &= e^{-\hat{T}}\hat{H} e^{\hat{T}} \nonumber\\
    &= \hat{H} + \left[\hat{H},\hat{T}\right] +{1\over 2!}\left[\left[\hat{H},\hat{T}\right],\hat{T}\right] +\ldots \ .
\end{align}
For two-body Hamiltonians, i.e., in the normal-ordered two-body approximation, this expansion truncates exactly at quadruply nested commutators. The matrix elements $\overline{H}_i^a$ and $\overline{H}_{ij}^{ab}$ of the Eqs.~(\ref{ccsd}) thus result from contracting the Hamiltonian matrix elements with those of the cluster amplitudes, i.e. lower (upper) indices in the Hamiltonian matrix elements match upper (lower) indices in the cluster amplitudes and are being summed over . Let us focus on the nonzero three-body matrix elements $W_{abc}^{cde}$ and $W_{ijk}^{lmn}$. The former have three incoming and three outgoing particle states, and contraction with the cluster amplitudes (which have incoming hole states and outgoing particle states) leaves the outgoing particle states $(cde)$ unchanged. Thus, they can only contribute to three-body matrix elements of $\overline{H}$ but not to the one- and two-body matrix elements of Eqs.~(\ref{ccsd}). Likewise, the matrix elements $W_{ijk}^{lmn}$ have three incoming and three outgoing hole states, and contraction with the cluster amplitudes leaves the incoming hole states $(ijk)$ unchanged. So these will also fail to contribute to the  one- and two-body matrix elements of Eqs.~(\ref{ccsd}). This completes the proof for the $^4$He nucleus.

It is now easy to extend the proof to other even-even $N=Z$ nuclei with mass number $A=4n$ where $n$ is a natural number. For the reference state we take a product state where $n$ lattice sites are each occupied by four nucleons and arranged in a suitable geometry for the corresponding ground state. This implies that the hole and particle states reside on different lattice sites. For $^{12}$C and $^{16}$O, examples are shown in Refs.~\cite{epelbaum2012,epelbaum2014}. Since these states also have nonzero overlap with the exact ground states, all arguments we used for $^4$He then carry through for these nuclei. 

The proof can also be extended to other nuclei. A lattice site belonging to the reference can hold up to four nucleons. If it is occupied by a single nucleon or by a pair of nucleons, an on-site three-body force has no effect on such configurations. If instead it is occupied by three nucleons, the three-nucleon force cannot flip the spin of the odd nucleon. Thus, our proof also holds here.  As a corollary we mention that the normal-ordered two-body truncation is also exact for coupled-cluster with singles computations.

\emph{Illustration of normal ordering truncation.}
We discovered the exact result when performing CCSD computations based on lattice Hamiltonians from pionless effective field theory at leading order. While our main statement does not need any numerical support, we would like to learn about the accuracy of the CCSD approximation for lattice Hamiltonians. 

The leading-order Hamiltonian from pionless effective field theory consists of the kinetic energy and zero-range (i.e., on-site) two- and three-body potentials~\cite{bedaque1999,bedaque2002}. It has been used in classical~\cite{platter2005,kirscher2010,lensky2016,barnea2015,kirscher2015,bansal2018,Konig:2019xxk} and quantum~\cite{roggero2020,baroni2022,gu2025} computations of light nuclei. 
\begin{align}
\label{hamLO}
    \hat{H} &= \sum_{\mathbf{l} \mathbf{l}'}\sum_{\tau s} T_{\mathbf{l}'}^{\mathbf{l}} \hat{a}_{\mathbf{l}\tau s}^\dagger \hat{a}_{\mathbf{l}'\tau s} \nonumber\\
    &+ {V\over 2}\sum_{\mathbf{l}}\sum_{ss'\tau\tau'} \hat{a}_{\mathbf{l}\tau s}^\dagger \hat{a}_{\mathbf{l}\tau' s'}^\dagger \hat{a}_{\mathbf{l}\tau' s'}\hat{a}_{\mathbf{l}\tau s} \\
    &+ W\sum_{\mathbf{l}}\sum_{\tau s} \hat{a}_{\mathbf{l}\tau\uparrow}^\dagger \hat{a}_{\mathbf{l}\tau\downarrow}^\dagger \hat{a}_{\mathbf{l}-\tau s}^\dagger \hat{a}_{\mathbf{l}-\tau s}\hat{a}_{\mathbf{l}\tau \downarrow}\hat{a}_{\mathbf{l}\tau \uparrow} \nonumber \ .
\end{align}
Here, the operator $\hat{a}^\dagger_{\mathbf{l}\tau s'}$ creates a nucleon on the lattice site $\mathbf{l}=(l_x, l_y, l_z)$ with the isospin projection $\tau$ and the spin projection $s$. The matrix elements of the kinetic energy are the leading-order approximation of the Laplacian on the lattice
\begin{align}
T_{\mathbf{l}'}^{\mathbf{l}} = -\frac{\hbar^2}{2ma^2}\sum_{i = x,y,z}\left(\delta_{\mathbf{l}'}^{\mathbf{l}-\mathbf{e}_i}    -2\delta_{\mathbf{l}'}^{\mathbf{l}} + \delta_{\mathbf{l}'}^{\mathbf{l}+\mathbf{e}_i} \right) .
\end{align}
Here $m$ is the nucleon mass, $a$ is the lattice spacing, and $\mathbf{e}_i$ is a unit vector in the direction $i=x,y,z$. The two-body and three-body contacts are
\begin{equation}
    V = \frac{\hbar^2v}{2ma^2} \quad\mbox{and}\quad W = \frac{\hbar^2w}{2ma^2} \ ,
\end{equation}
respectively. Here $v$ and $w$ are dimensionless couplings. 
The Hamiltonian~(\ref{hamLO}) is spin-isospin invariant and exhibits Wigner's SU(4) symmetry.

For three different lattice spacings $a$ we adjusted the parameters $v$ and $w$ such that on lattices with $L=4$ we obtained ground-state energies for $^2$H and $^4$He that are consistent within pionless effective field theory (where an error of about 1/3 is acceptable). Results are shown in Table~\ref{tab:results}. We did not apply any finite size corrections and remind the reader that the $L\to\infty$ results are above those we obtained~\cite{luscher1985,konig2017}. We limited ourselves to $L=4$, because the Hamiltonian of the $^4$He nucleus then has a matrix dimension of $(L^3)^4\approx 17\times 10^6$ and its ground state can still be computed on a laptop.  

\begin{table}
\renewcommand{\arraystretch}{1.2}
\centering
\caption{Ground-state energies of $^2$H and $^{3,4}$He (in MeV) from exact diagonalizations and from coupled-cluster with singles and doubles (CCSD) on lattices of extent $L=4$ and a lattice spacing $a$ (in fm).  The two-body coupling $v$ and three-body coupling $w$ were adjusted to the ground-state energies of $^2$H and $^4$He.}
\label{tab:results}
\renewcommand{\arraystretch}{1.3}
\begin{ruledtabular}
\begin{tabular}{cccrrrrr}
\multicolumn{3}{c}{Parameters} & \multicolumn{2}{c}{$^4$He} & \multicolumn{2}{c}{$^3$He} & $^2$H\\\cline{1-3}\cline{4-5}\cline{6-7}
$a$ & $v$ & $w$ & $E_{\rm exact}$ & $E_{\rm CCSD}$ & $E_{\rm exact}$ & $E_{\rm CCSD}$ &$E_{\rm exact}$\\ 
 \colrule
2.5 & $-9.0$ & 6.0 & $-29.70$ & $-29.11$ & $-16.32$ & $-15.22$ & $-2.48$\\
2.0 & $-8.0$ & 5.5 & $-29.45$ & $-27.73$ & $-14.84$ & $-12.18$ & $-2.53$\\
1.7 & $-7.0$ & 4.4 & $-28.97$ & $-26.41$ & $-10.82$ & $-4.80$ & $-2.33$
\end{tabular}
\end{ruledtabular}
\end{table}

For the CCSD computations of the $A=3$ and 4 nuclei we use reference states where all nucleons are on a single lattice site. The energy of the reference state then is
\begin{equation}
\label{Eref}
    E_{\rm ref} = \frac{\hbar^2}{2ma^2}\left[6A +\binom{A}{2}v +\binom{A}{3}w \right] \ .
\end{equation}

\begin{figure}[t!]
    \centering
    \includegraphics[width=0.49\textwidth]{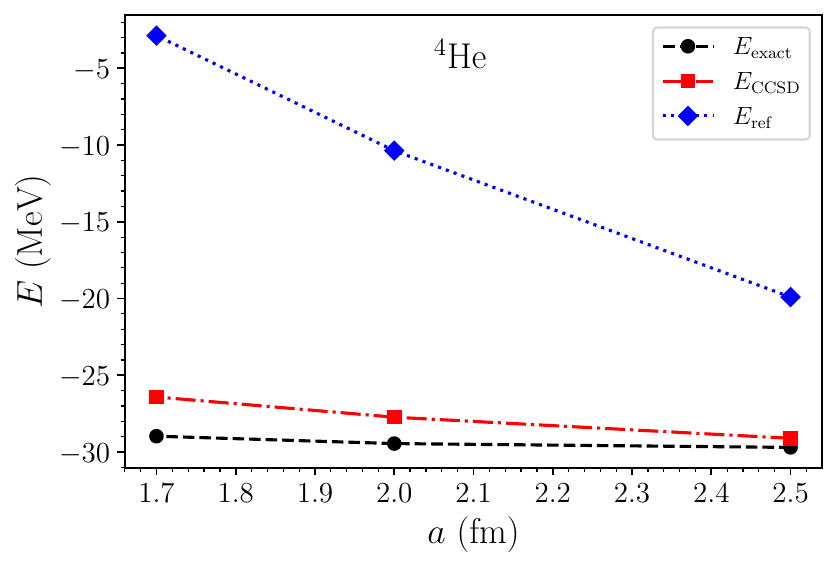}
    \caption{Contributions to the ground-state energy of $^4$He as a function of the lattice spacing $a$. Energy $E_{\rm ref}$ of the reference state (blue diamonds connected by dotted line), CCSD energy $E_{\rm CCSD}$ (red squares connected by dashed-dotted line) and the exact energy $E_{\rm exact}$ (black circles connected by dashed lines). }
    \label{fig:4He}
\end{figure}

The coupled-cluster results for the ground-state energies of $^{3,4}$He are also shown in Table~\ref{tab:results}. The correlation energy is defined as the difference between the ground-state energy and the energy~(\ref{Eref}) of the reference state. We remind the reader that CCSD usually captures about 90\% of the correlation energy in quantum chemistry and in nuclear physics applications for systems with a well defined gap at the Fermi surface~\cite{bartlett2007,hagen2009b,sun2022,sun2025}; it can be even more in light nuclei~\cite{hagen2007b,roth2012}.
Figure~\ref{fig:4He} shows the energy contributions to $^4$He as a function of the lattice spacing $a$. Inspection shows that the CCSD energy $E_{\rm CCSD}$ captures about 90\% of the correlation energy.  

\begin{figure}[t!]
    \centering
    \includegraphics[width=0.49\textwidth]{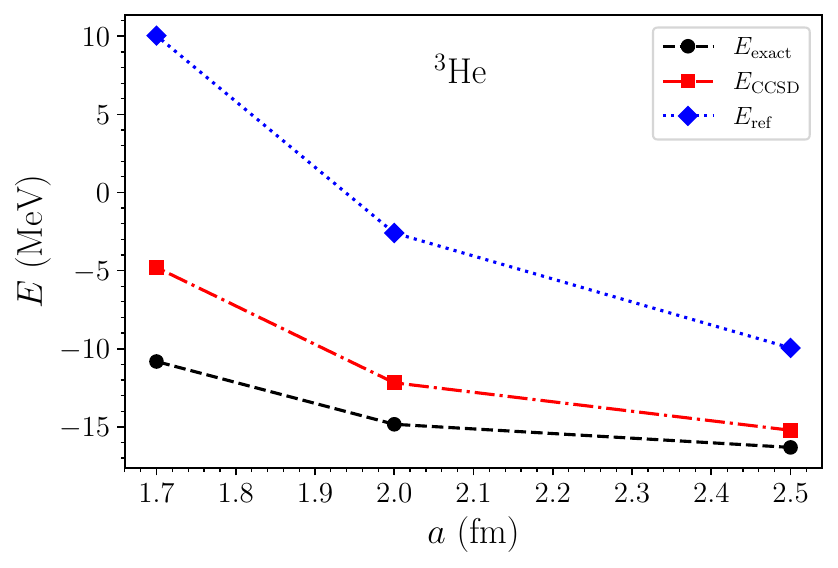}
    \caption{Same as Fig.~\ref{fig:4He} but for $^3$He.}
    \label{fig:3He}
\end{figure}

The situation is a little bit different for $^3$He, and results are shown in Fig.~\ref{fig:3He}. We find that CCSD yields 83, 78 and 71\% of the correlation energy for lattice spacings $a=2.5$, 2.0, and 1.7~fm, respectively. We observe a similar behavior when there is no three-nucleon force; there CCSD yields 87, 85, and 82\%
for $a=2.5$, 2.0, and 1.7~fm, respectively. So it seems that the structure of this nucleus---with or without three nucleon forces---is more challenging to capture with the CCSD method on the lattice.

\emph{Discussion.}
Let us discuss the results of this Letter in the context of (i) finite-range forces, (ii) three-particle--three-hole (triples) excitations, and (iii) the extension to four-body forces.

First, finite range forces, i.e., from pion exchange, have a range proportional to the inverse pion mass $m_\pi^{-1}\approx 1.4$~fm and fall off exponentially fast with that length scale. Thus, they are smaller corrections when compared to the on-site contributions. This explains why the comparisons of the normal-ordered two-body truncation with exact results in Refs.~\cite{hagen2007a,roth2012} were so accurate.

Second, if one extends the computational method to  coupled-cluster theory with singles, doubles, and triples, the nonzero three-body matrix elements $W_{abc}^{cde}$ and $W_{ijk}^{lmn}$ do contribute to the solution of the coupled-cluster equations for the triples amplitudes. However, as discussed above $W_{abc}^{ijk}=0$ and triples amplitudes do not yield a direct contribution to the correlation energy. Instead, the impact of residual three-nucleon forces onto the energy is only indirect, i.e., by changing the singles and doubles amplitudes which then contribute to the correlation energy. This is also reflective of the fact that the leading contributions of the residual three-nucleon force first appear at seventh order in perturbation theory in this case. This is so because it takes three insertions of the kinetic energy to move three particles to a neighboring site, followed by one three-body interaction  and subsequent three insertions of the kinetic energy to move the three nucleons back to the original site. If the two-body force had a finite range, the leading contributions of the residual three-nucleon force would first appear at fifth order in perturbation theory (because two insertions of the kinetic energy could be replaced by a single insertion of the two-body potential). Nevertheless, the case of $^3$He shows that this could give significant  contributions to the correlation energy. 

Third, let us discuss four-nucleon forces. They enter in pionless effective field theory at next-to-leading order~\cite{bazak2019} and in chiral effective field theory at next-to-next-to-next-to-leading order in the Weinberg power counting.  Again, we might take the four-body contact as a zero-range (i.e., on-site) interaction. It is then again  clear that only $U_{abcd}^{efgh}$ and $U_{ijki'}^{lmnl'}$ are not all vanishing. The matrix elements $U_{ijki'}^{lmnl'}$ contribute to the operators of the normal-ordered Hamiltonian with ranks up to three, but the nonzero four-body matrix elements can only impact the coupled-cluster amplitudes once four-particle–four-hole excitations (quadruples) are included.

\emph{Summary.} The normal-ordered two-body approximation is a key approximation in applications of \emph{ab initio} methods that target medium-mass and heavy nuclei. This approximation truncates a nuclear Hamiltonian (which includes three-nucleon forces) at the two-body level after normal ordering with respect to a nontrivial reference state has taken place.  We showed that this truncation is exact for zero-ranged three-nucleon interactions when computations use the coupled-cluster with singles and doubles method. Corrections then come from finite ranges of the three-nucleon forces and from three-particle--three-hole excitations. The analytical result lends credibility to the popular normal-ordered two-body approximation. It is rare to obtain analytical insights into the nuclear many-body problem. This is one of those cases. 

\emph{Acknowledgments.}
This work was supported by the U.S. Department of Energy, Office of
Science, Office of Nuclear Physics, under Award No.~DE-FG02-96ER40963 and under the FRIB
Theory Alliance award DE-SC0013617; by the U.S. Department of Energy, Office of Science, Office of Advanced Scientific Computing Research and Office of Nuclear Physics, Scientific Discovery through Advanced Computing (SciDAC) program (SciDAC-5 NUCLEI); and by the Laboratory Directed Research and Development Program of Oak Ridge National Laboratory, managed by UT-Battelle, LLC, for the U.S.\ Department of Energy. Oak Ridge National Laboratory is
supported by the Office of Science of the Department of Energy under
contract No. DE-AC05-00OR22725.
This research used resources of the Oak Ridge Leadership Computing Facility located at Oak Ridge National Laboratory. 

This manuscript has been authored in part by UT-Battelle, LLC, under contract DE-AC05-00OR22725 with the US Department of Energy (DOE). The US government retains and the publisher, by accepting the article for publication, acknowledges that the US government retains a nonexclusive, paid-up, irrevocable, worldwide license to publish or reproduce the published form of this manuscript, or allow others to do so, for US government purposes. DOE will provide public access to these results of federally sponsored research in accordance with the DOE Public Access Plan (\url{http://energy.gov/downloads/doe-public-access-plan}).

\emph{Data availability.}
The data that support the findings of this article are openly available~\cite{rothman_zenodo2025}. 

\bibliography{master-merged}

\begin{thebibliography}{60}%
\makeatletter
\providecommand \@ifxundefined [1]{%
 \@ifx{#1\undefined}
}%
\providecommand \@ifnum [1]{%
 \ifnum #1\expandafter \@firstoftwo
 \else \expandafter \@secondoftwo
 \fi
}%
\providecommand \@ifx [1]{%
 \ifx #1\expandafter \@firstoftwo
 \else \expandafter \@secondoftwo
 \fi
}%
\providecommand \natexlab [1]{#1}%
\providecommand \enquote  [1]{``#1''}%
\providecommand \bibnamefont  [1]{#1}%
\providecommand \bibfnamefont [1]{#1}%
\providecommand \citenamefont [1]{#1}%
\providecommand \href@noop [0]{\@secondoftwo}%
\providecommand \href [0]{\begingroup \@sanitize@url \@href}%
\providecommand \@href[1]{\@@startlink{#1}\@@href}%
\providecommand \@@href[1]{\endgroup#1\@@endlink}%
\providecommand \@sanitize@url [0]{\catcode `\\12\catcode `\$12\catcode
  `\&12\catcode `\#12\catcode `\^12\catcode `\_12\catcode `\%12\relax}%
\providecommand \@@startlink[1]{}%
\providecommand \@@endlink[0]{}%
\providecommand \url  [0]{\begingroup\@sanitize@url \@url }%
\providecommand \@url [1]{\endgroup\@href {#1}{\urlprefix }}%
\providecommand \urlprefix  [0]{URL }%
\providecommand \Eprint [0]{\href }%
\providecommand \doibase [0]{https://doi.org/}%
\providecommand \selectlanguage [0]{\@gobble}%
\providecommand \bibinfo  [0]{\@secondoftwo}%
\providecommand \bibfield  [0]{\@secondoftwo}%
\providecommand \translation [1]{[#1]}%
\providecommand \BibitemOpen [0]{}%
\providecommand \bibitemStop [0]{}%
\providecommand \bibitemNoStop [0]{.\EOS\space}%
\providecommand \EOS [0]{\spacefactor3000\relax}%
\providecommand \BibitemShut  [1]{\csname bibitem#1\endcsname}%
\let\auto@bib@innerbib\@empty
\bibitem [{\citenamefont {Hu}\ \emph {et~al.}(2022)\citenamefont {Hu},
  \citenamefont {Jiang}, \citenamefont {Miyagi}, \citenamefont {Sun},
  \citenamefont {Ekstr{\"o}m}, \citenamefont {Forss{\'e}n}, \citenamefont
  {Hagen}, \citenamefont {Holt}, \citenamefont {Papenbrock}, \citenamefont
  {Stroberg},\ and\ \citenamefont {Vernon}}]{hu2022}%
  \BibitemOpen
  \bibfield  {author} {\bibinfo {author} {\bibfnamefont {B.}~\bibnamefont
  {Hu}}, \bibinfo {author} {\bibfnamefont {W.}~\bibnamefont {Jiang}}, \bibinfo
  {author} {\bibfnamefont {T.}~\bibnamefont {Miyagi}}, \bibinfo {author}
  {\bibfnamefont {Z.}~\bibnamefont {Sun}}, \bibinfo {author} {\bibfnamefont
  {A.}~\bibnamefont {Ekstr{\"o}m}}, \bibinfo {author} {\bibfnamefont
  {C.}~\bibnamefont {Forss{\'e}n}}, \bibinfo {author} {\bibfnamefont
  {G.}~\bibnamefont {Hagen}}, \bibinfo {author} {\bibfnamefont {J.~D.}\
  \bibnamefont {Holt}}, \bibinfo {author} {\bibfnamefont {T.}~\bibnamefont
  {Papenbrock}}, \bibinfo {author} {\bibfnamefont {S.~R.}\ \bibnamefont
  {Stroberg}},\ and\ \bibinfo {author} {\bibfnamefont {I.}~\bibnamefont
  {Vernon}},\ }\bibfield  {title} {\bibinfo {title} {Ab initio predictions link
  the neutron skin of $^{208}\mathrm{Pb}$ to nuclear forces},\ }\href
  {https://doi.org/10.1038/s41567-022-01715-8} {\bibfield  {journal} {\bibinfo
  {journal} {Nat. Phys.}\ }\textbf {\bibinfo {volume} {18}},\ \bibinfo {pages}
  {1196} (\bibinfo {year} {2022})}\BibitemShut {NoStop}%
\bibitem [{\citenamefont {Hebeler}\ \emph {et~al.}(2023)\citenamefont
  {Hebeler}, \citenamefont {Durant}, \citenamefont {Hoppe}, \citenamefont
  {Heinz}, \citenamefont {Schwenk}, \citenamefont {Simonis},\ and\
  \citenamefont {Tichai}}]{hebeler2023}%
  \BibitemOpen
  \bibfield  {author} {\bibinfo {author} {\bibfnamefont {K.}~\bibnamefont
  {Hebeler}}, \bibinfo {author} {\bibfnamefont {V.}~\bibnamefont {Durant}},
  \bibinfo {author} {\bibfnamefont {J.}~\bibnamefont {Hoppe}}, \bibinfo
  {author} {\bibfnamefont {M.}~\bibnamefont {Heinz}}, \bibinfo {author}
  {\bibfnamefont {A.}~\bibnamefont {Schwenk}}, \bibinfo {author} {\bibfnamefont
  {J.}~\bibnamefont {Simonis}},\ and\ \bibinfo {author} {\bibfnamefont
  {A.}~\bibnamefont {Tichai}},\ }\bibfield  {title} {\bibinfo {title} {Normal
  ordering of three-nucleon interactions for ab initio calculations of heavy
  nuclei},\ }\href {https://doi.org/10.1103/PhysRevC.107.024310} {\bibfield
  {journal} {\bibinfo  {journal} {Phys. Rev. C}\ }\textbf {\bibinfo {volume}
  {107}},\ \bibinfo {pages} {024310} (\bibinfo {year} {2023})}\BibitemShut
  {NoStop}%
\bibitem [{\citenamefont {Miyagi}\ \emph {et~al.}(2024)\citenamefont {Miyagi},
  \citenamefont {Cao}, \citenamefont {Seutin}, \citenamefont {Bacca},
  \citenamefont {Garcia~Ruiz}, \citenamefont {Hebeler}, \citenamefont {Holt},\
  and\ \citenamefont {Schwenk}}]{miyagi2024}%
  \BibitemOpen
  \bibfield  {author} {\bibinfo {author} {\bibfnamefont {T.}~\bibnamefont
  {Miyagi}}, \bibinfo {author} {\bibfnamefont {X.}~\bibnamefont {Cao}},
  \bibinfo {author} {\bibfnamefont {R.}~\bibnamefont {Seutin}}, \bibinfo
  {author} {\bibfnamefont {S.}~\bibnamefont {Bacca}}, \bibinfo {author}
  {\bibfnamefont {R.~F.}\ \bibnamefont {Garcia~Ruiz}}, \bibinfo {author}
  {\bibfnamefont {K.}~\bibnamefont {Hebeler}}, \bibinfo {author} {\bibfnamefont
  {J.~D.}\ \bibnamefont {Holt}},\ and\ \bibinfo {author} {\bibfnamefont
  {A.}~\bibnamefont {Schwenk}},\ }\bibfield  {title} {\bibinfo {title} {Impact
  of two-body currents on magnetic dipole moments of nuclei},\ }\href
  {https://doi.org/10.1103/physrevlett.132.232503} {\bibfield  {journal}
  {\bibinfo  {journal} {Phys. Rev. Lett.}\ }\textbf {\bibinfo {volume} {132}},\
  \bibinfo {pages} {232503} (\bibinfo {year} {2024})}\BibitemShut {NoStop}%
\bibitem [{\citenamefont {Hergert}(2020)}]{hergert2020}%
  \BibitemOpen
  \bibfield  {author} {\bibinfo {author} {\bibfnamefont {H.}~\bibnamefont
  {Hergert}},\ }\bibfield  {title} {\bibinfo {title} {A guided tour of ab
  initio nuclear many-body theory},\ }\href
  {https://doi.org/10.3389/fphy.2020.00379} {\bibfield  {journal} {\bibinfo
  {journal} {Front. Phys.}\ }\textbf {\bibinfo {volume} {8}},\ \bibinfo {pages}
  {379} (\bibinfo {year} {2020})}\BibitemShut {NoStop}%
\bibitem [{\citenamefont {Ekstr\"om}\ \emph {et~al.}(2023)\citenamefont
  {Ekstr\"om}, \citenamefont {Forss\'en}, \citenamefont {Hagen}, \citenamefont
  {Jansen}, \citenamefont {Jiang},\ and\ \citenamefont
  {Papenbrock}}]{Ekstrom:2022yea}%
  \BibitemOpen
  \bibfield  {author} {\bibinfo {author} {\bibfnamefont {A.}~\bibnamefont
  {Ekstr\"om}}, \bibinfo {author} {\bibfnamefont {C.}~\bibnamefont
  {Forss\'en}}, \bibinfo {author} {\bibfnamefont {G.}~\bibnamefont {Hagen}},
  \bibinfo {author} {\bibfnamefont {G.~R.}\ \bibnamefont {Jansen}}, \bibinfo
  {author} {\bibfnamefont {W.}~\bibnamefont {Jiang}},\ and\ \bibinfo {author}
  {\bibfnamefont {T.}~\bibnamefont {Papenbrock}},\ }\bibfield  {title}
  {\bibinfo {title} {{What is ab initio in nuclear theory?}},\ }\href
  {https://doi.org/10.3389/fphy.2023.1129094} {\bibfield  {journal} {\bibinfo
  {journal} {Front. Phys.}\ }\textbf {\bibinfo {volume} {11}},\ \bibinfo
  {pages} {1129094} (\bibinfo {year} {2023})}\BibitemShut {NoStop}%
\bibitem [{\citenamefont {Bedaque}\ \emph {et~al.}(1999)\citenamefont
  {Bedaque}, \citenamefont {Hammer},\ and\ \citenamefont {van
  Kolck}}]{bedaque1999}%
  \BibitemOpen
  \bibfield  {author} {\bibinfo {author} {\bibfnamefont {P.~F.}\ \bibnamefont
  {Bedaque}}, \bibinfo {author} {\bibfnamefont {H.-W.}\ \bibnamefont
  {Hammer}},\ and\ \bibinfo {author} {\bibfnamefont {U.}~\bibnamefont {van
  Kolck}},\ }\bibfield  {title} {\bibinfo {title} {Renormalization of the
  three-body system with short-range interactions},\ }\href
  {https://doi.org/10.1103/PhysRevLett.82.463} {\bibfield  {journal} {\bibinfo
  {journal} {Phys. Rev. Lett.}\ }\textbf {\bibinfo {volume} {82}},\ \bibinfo
  {pages} {463} (\bibinfo {year} {1999})}\BibitemShut {NoStop}%
\bibitem [{\citenamefont {{Bedaque}}\ and\ \citenamefont {{van
  Kolck}}(2002)}]{bedaque2002}%
  \BibitemOpen
  \bibfield  {author} {\bibinfo {author} {\bibfnamefont {P.~F.}\ \bibnamefont
  {{Bedaque}}}\ and\ \bibinfo {author} {\bibfnamefont {U.}~\bibnamefont {{van
  Kolck}}},\ }\bibfield  {title} {\bibinfo {title} {{Effective field theory for
  few-nucleon systems}},\ }\href
  {https://doi.org/10.1146/annurev.nucl.52.050102.090637} {\bibfield  {journal}
  {\bibinfo  {journal} {Ann. Rev. Nucl. Part. Sci.}\ }\textbf {\bibinfo
  {volume} {52}},\ \bibinfo {pages} {339} (\bibinfo {year} {2002})}\BibitemShut
  {NoStop}%
\bibitem [{\citenamefont {Platter}\ \emph {et~al.}(2005)\citenamefont
  {Platter}, \citenamefont {Hammer},\ and\ \citenamefont
  {Mei{\ss}ner}}]{platter2005}%
  \BibitemOpen
  \bibfield  {author} {\bibinfo {author} {\bibfnamefont {L.}~\bibnamefont
  {Platter}}, \bibinfo {author} {\bibfnamefont {H.-W.}\ \bibnamefont
  {Hammer}},\ and\ \bibinfo {author} {\bibfnamefont {U.-G.}\ \bibnamefont
  {Mei{\ss}ner}},\ }\bibfield  {title} {\bibinfo {title} {On the correlation
  between the binding energies of the triton and the $\alpha$-particle},\
  }\href {https://doi.org/10.1016/j.physletb.2004.12.068} {\bibfield  {journal}
  {\bibinfo  {journal} {Phys. Lett. B}\ }\textbf {\bibinfo {volume} {607}},\
  \bibinfo {pages} {254 } (\bibinfo {year} {2005})}\BibitemShut {NoStop}%
\bibitem [{\citenamefont {Kirscher}\ \emph {et~al.}(2010)\citenamefont
  {Kirscher}, \citenamefont {Grie{\ss}hammer}, \citenamefont {Shukla},\ and\
  \citenamefont {Hofmann}}]{kirscher2010}%
  \BibitemOpen
  \bibfield  {author} {\bibinfo {author} {\bibfnamefont {J.}~\bibnamefont
  {Kirscher}}, \bibinfo {author} {\bibfnamefont {H.~W.}\ \bibnamefont
  {Grie{\ss}hammer}}, \bibinfo {author} {\bibfnamefont {D.}~\bibnamefont
  {Shukla}},\ and\ \bibinfo {author} {\bibfnamefont {H.~M.}\ \bibnamefont
  {Hofmann}},\ }\bibfield  {title} {\bibinfo {title} {Universal correlations in
  pion-less {EFT} with the resonating group method: Three and four nucleons},\
  }\href {https://doi.org/10.1140/epja/i2010-10939-5} {\bibfield  {journal}
  {\bibinfo  {journal} {Eur. Phys. J. A}\ }\textbf {\bibinfo {volume} {44}},\
  \bibinfo {pages} {239} (\bibinfo {year} {2010})}\BibitemShut {NoStop}%
\bibitem [{\citenamefont {Barnea}\ \emph {et~al.}(2015)\citenamefont {Barnea},
  \citenamefont {Contessi}, \citenamefont {Gazit}, \citenamefont {Pederiva},\
  and\ \citenamefont {van Kolck}}]{barnea2015}%
  \BibitemOpen
  \bibfield  {author} {\bibinfo {author} {\bibfnamefont {N.}~\bibnamefont
  {Barnea}}, \bibinfo {author} {\bibfnamefont {L.}~\bibnamefont {Contessi}},
  \bibinfo {author} {\bibfnamefont {D.}~\bibnamefont {Gazit}}, \bibinfo
  {author} {\bibfnamefont {F.}~\bibnamefont {Pederiva}},\ and\ \bibinfo
  {author} {\bibfnamefont {U.}~\bibnamefont {van Kolck}},\ }\bibfield  {title}
  {\bibinfo {title} {Effective field theory for lattice nuclei},\ }\href
  {https://doi.org/10.1103/PhysRevLett.114.052501} {\bibfield  {journal}
  {\bibinfo  {journal} {Phys. Rev. Lett.}\ }\textbf {\bibinfo {volume} {114}},\
  \bibinfo {pages} {052501} (\bibinfo {year} {2015})}\BibitemShut {NoStop}%
\bibitem [{\citenamefont {Kirscher}\ \emph {et~al.}(2015)\citenamefont
  {Kirscher}, \citenamefont {Barnea}, \citenamefont {Gazit}, \citenamefont
  {Pederiva},\ and\ \citenamefont {van Kolck}}]{kirscher2015}%
  \BibitemOpen
  \bibfield  {author} {\bibinfo {author} {\bibfnamefont {J.}~\bibnamefont
  {Kirscher}}, \bibinfo {author} {\bibfnamefont {N.}~\bibnamefont {Barnea}},
  \bibinfo {author} {\bibfnamefont {D.}~\bibnamefont {Gazit}}, \bibinfo
  {author} {\bibfnamefont {F.}~\bibnamefont {Pederiva}},\ and\ \bibinfo
  {author} {\bibfnamefont {U.}~\bibnamefont {van Kolck}},\ }\bibfield  {title}
  {\bibinfo {title} {Spectra and scattering of light lattice nuclei from
  effective field theory},\ }\href {https://doi.org/10.1103/PhysRevC.92.054002}
  {\bibfield  {journal} {\bibinfo  {journal} {Phys. Rev. C}\ }\textbf {\bibinfo
  {volume} {92}},\ \bibinfo {pages} {054002} (\bibinfo {year}
  {2015})}\BibitemShut {NoStop}%
\bibitem [{\citenamefont {Lensky}\ \emph {et~al.}(2016)\citenamefont {Lensky},
  \citenamefont {Birse},\ and\ \citenamefont {Walet}}]{lensky2016}%
  \BibitemOpen
  \bibfield  {author} {\bibinfo {author} {\bibfnamefont {V.}~\bibnamefont
  {Lensky}}, \bibinfo {author} {\bibfnamefont {M.~C.}\ \bibnamefont {Birse}},\
  and\ \bibinfo {author} {\bibfnamefont {N.~R.}\ \bibnamefont {Walet}},\
  }\bibfield  {title} {\bibinfo {title} {Description of light nuclei in
  pionless effective field theory using the stochastic variational method},\
  }\href {https://doi.org/10.1103/PhysRevC.94.034003} {\bibfield  {journal}
  {\bibinfo  {journal} {Phys. Rev. C}\ }\textbf {\bibinfo {volume} {94}},\
  \bibinfo {pages} {034003} (\bibinfo {year} {2016})}\BibitemShut {NoStop}%
\bibitem [{\citenamefont {Bansal}\ \emph {et~al.}(2018)\citenamefont {Bansal},
  \citenamefont {Binder}, \citenamefont {Ekstr\"om}, \citenamefont {Hagen},
  \citenamefont {Jansen},\ and\ \citenamefont {Papenbrock}}]{bansal2018}%
  \BibitemOpen
  \bibfield  {author} {\bibinfo {author} {\bibfnamefont {A.}~\bibnamefont
  {Bansal}}, \bibinfo {author} {\bibfnamefont {S.}~\bibnamefont {Binder}},
  \bibinfo {author} {\bibfnamefont {A.}~\bibnamefont {Ekstr\"om}}, \bibinfo
  {author} {\bibfnamefont {G.}~\bibnamefont {Hagen}}, \bibinfo {author}
  {\bibfnamefont {G.~R.}\ \bibnamefont {Jansen}},\ and\ \bibinfo {author}
  {\bibfnamefont {T.}~\bibnamefont {Papenbrock}},\ }\bibfield  {title}
  {\bibinfo {title} {Pion-less effective field theory for atomic nuclei and
  lattice nuclei},\ }\href {https://doi.org/10.1103/PhysRevC.98.054301}
  {\bibfield  {journal} {\bibinfo  {journal} {Phys. Rev. C}\ }\textbf {\bibinfo
  {volume} {98}},\ \bibinfo {pages} {054301} (\bibinfo {year}
  {2018})}\BibitemShut {NoStop}%
\bibitem [{\citenamefont {K{\"o}nig}(2020)}]{Konig:2019xxk}%
  \BibitemOpen
  \bibfield  {author} {\bibinfo {author} {\bibfnamefont {S.}~\bibnamefont
  {K{\"o}nig}},\ }\bibfield  {title} {\bibinfo {title} {{Energies and radii of
  light nuclei around unitarity}},\ }\href
  {https://doi.org/10.1140/epja/s10050-020-00098-9} {\bibfield  {journal}
  {\bibinfo  {journal} {Eur. Phys. J. A}\ }\textbf {\bibinfo {volume} {56}},\
  \bibinfo {pages} {113} (\bibinfo {year} {2020})}\BibitemShut {NoStop}%
\bibitem [{\citenamefont {Epelbaum}(2006)}]{epelbaum06}%
  \BibitemOpen
  \bibfield  {author} {\bibinfo {author} {\bibfnamefont {E.}~\bibnamefont
  {Epelbaum}},\ }\bibfield  {title} {\bibinfo {title} {Few-nucleon forces and
  systems in chiral effective field theory},\ }\href
  {https://doi.org/10.1007/s00601-008-0209-7} {\bibfield  {journal} {\bibinfo
  {journal} {Prog. Part. Nucl. Phys.}\ }\textbf {\bibinfo {volume} {57}},\
  \bibinfo {pages} {654 } (\bibinfo {year} {2006})}\BibitemShut {NoStop}%
\bibitem [{\citenamefont {Epelbaum}\ \emph {et~al.}(2009)\citenamefont
  {Epelbaum}, \citenamefont {Hammer},\ and\ \citenamefont
  {Mei\ss{}ner}}]{epelbaum2009}%
  \BibitemOpen
  \bibfield  {author} {\bibinfo {author} {\bibfnamefont {E.}~\bibnamefont
  {Epelbaum}}, \bibinfo {author} {\bibfnamefont {H.-W.}\ \bibnamefont
  {Hammer}},\ and\ \bibinfo {author} {\bibfnamefont {U.-G.}\ \bibnamefont
  {Mei\ss{}ner}},\ }\bibfield  {title} {\bibinfo {title} {Modern theory of
  nuclear forces},\ }\href {https://doi.org/10.1103/RevModPhys.81.1773}
  {\bibfield  {journal} {\bibinfo  {journal} {Rev. Mod. Phys.}\ }\textbf
  {\bibinfo {volume} {81}},\ \bibinfo {pages} {1773} (\bibinfo {year}
  {2009})}\BibitemShut {NoStop}%
\bibitem [{\citenamefont {Machleidt}\ and\ \citenamefont
  {Entem}(2011)}]{machleidt2011}%
  \BibitemOpen
  \bibfield  {author} {\bibinfo {author} {\bibfnamefont {R.}~\bibnamefont
  {Machleidt}}\ and\ \bibinfo {author} {\bibfnamefont {D.}~\bibnamefont
  {Entem}},\ }\bibfield  {title} {\bibinfo {title} {Chiral effective field
  theory and nuclear forces},\ }\href
  {https://doi.org/10.1016/j.physrep.2011.02.001} {\bibfield  {journal}
  {\bibinfo  {journal} {Phys. Rep.}\ }\textbf {\bibinfo {volume} {503}},\
  \bibinfo {pages} {1 } (\bibinfo {year} {2011})}\BibitemShut {NoStop}%
\bibitem [{\citenamefont {Ekstr\"om}\ \emph {et~al.}(2015)\citenamefont
  {Ekstr\"om}, \citenamefont {Jansen}, \citenamefont {Wendt}, \citenamefont
  {Hagen}, \citenamefont {Papenbrock}, \citenamefont {Carlsson}, \citenamefont
  {Forss\'en}, \citenamefont {Hjorth-Jensen}, \citenamefont {Navr\'atil},\ and\
  \citenamefont {Nazarewicz}}]{ekstrom2015a}%
  \BibitemOpen
  \bibfield  {author} {\bibinfo {author} {\bibfnamefont {A.}~\bibnamefont
  {Ekstr\"om}}, \bibinfo {author} {\bibfnamefont {G.~R.}\ \bibnamefont
  {Jansen}}, \bibinfo {author} {\bibfnamefont {K.~A.}\ \bibnamefont {Wendt}},
  \bibinfo {author} {\bibfnamefont {G.}~\bibnamefont {Hagen}}, \bibinfo
  {author} {\bibfnamefont {T.}~\bibnamefont {Papenbrock}}, \bibinfo {author}
  {\bibfnamefont {B.~D.}\ \bibnamefont {Carlsson}}, \bibinfo {author}
  {\bibfnamefont {C.}~\bibnamefont {Forss\'en}}, \bibinfo {author}
  {\bibfnamefont {M.}~\bibnamefont {Hjorth-Jensen}}, \bibinfo {author}
  {\bibfnamefont {P.}~\bibnamefont {Navr\'atil}},\ and\ \bibinfo {author}
  {\bibfnamefont {W.}~\bibnamefont {Nazarewicz}},\ }\bibfield  {title}
  {\bibinfo {title} {Accurate nuclear radii and binding energies from a chiral
  interaction},\ }\href {https://doi.org/10.1103/PhysRevC.91.051301} {\bibfield
   {journal} {\bibinfo  {journal} {Phys. Rev. C}\ }\textbf {\bibinfo {volume}
  {91}},\ \bibinfo {pages} {051301} (\bibinfo {year} {2015})}\BibitemShut
  {NoStop}%
\bibitem [{\citenamefont {Reinert}\ \emph {et~al.}(2018)\citenamefont
  {Reinert}, \citenamefont {Krebs},\ and\ \citenamefont
  {Epelbaum}}]{reinert2018}%
  \BibitemOpen
  \bibfield  {author} {\bibinfo {author} {\bibfnamefont {P.}~\bibnamefont
  {Reinert}}, \bibinfo {author} {\bibfnamefont {H.}~\bibnamefont {Krebs}},\
  and\ \bibinfo {author} {\bibfnamefont {E.}~\bibnamefont {Epelbaum}},\
  }\bibfield  {title} {\bibinfo {title} {Semilocal momentum-space regularized
  chiral two-nucleon potentials up to fifth order},\ }\href
  {https://doi.org/10.1140/epja/i2018-12516-4} {\bibfield  {journal} {\bibinfo
  {journal} {Eur. Phys. J. A}\ }\textbf {\bibinfo {volume} {54}},\ \bibinfo
  {pages} {86} (\bibinfo {year} {2018})}\BibitemShut {NoStop}%
\bibitem [{\citenamefont {Piarulli}\ and\ \citenamefont
  {Tews}(2020)}]{Piarulli:2019cqu}%
  \BibitemOpen
  \bibfield  {author} {\bibinfo {author} {\bibfnamefont {M.}~\bibnamefont
  {Piarulli}}\ and\ \bibinfo {author} {\bibfnamefont {I.}~\bibnamefont
  {Tews}},\ }\bibfield  {title} {\bibinfo {title} {{Local Nucleon-Nucleon and
  Three-Nucleon Interactions Within Chiral Effective Field Theory}},\ }\href
  {https://doi.org/10.3389/fphy.2019.00245} {\bibfield  {journal} {\bibinfo
  {journal} {Front. Phys.}\ }\textbf {\bibinfo {volume} {7}},\ \bibinfo {pages}
  {245} (\bibinfo {year} {2020})}\BibitemShut {NoStop}%
\bibitem [{\citenamefont {Dickhoff}\ and\ \citenamefont
  {Barbieri}(2004)}]{dickhoff2004}%
  \BibitemOpen
  \bibfield  {author} {\bibinfo {author} {\bibfnamefont {W.}~\bibnamefont
  {Dickhoff}}\ and\ \bibinfo {author} {\bibfnamefont {C.}~\bibnamefont
  {Barbieri}},\ }\bibfield  {title} {\bibinfo {title} {Self-consistent
  {G}reen's function method for nuclei and nuclear matter},\ }\href
  {https://doi.org/10.1016/j.ppnp.2004.02.038} {\bibfield  {journal} {\bibinfo
  {journal} {Prog. Part. Nucl. Phys.}\ }\textbf {\bibinfo {volume} {52}},\
  \bibinfo {pages} {377 } (\bibinfo {year} {2004})}\BibitemShut {NoStop}%
\bibitem [{\citenamefont {Som\`a}\ and\ \citenamefont {Bo\ifmmode~\dot{z}\else
  \.{z}\fi{}ek}(2008)}]{soma2008}%
  \BibitemOpen
  \bibfield  {author} {\bibinfo {author} {\bibfnamefont {V.}~\bibnamefont
  {Som\`a}}\ and\ \bibinfo {author} {\bibfnamefont {P.}~\bibnamefont
  {Bo\ifmmode~\dot{z}\else \.{z}\fi{}ek}},\ }\bibfield  {title} {\bibinfo
  {title} {In-medium {$T$} matrix for nuclear matter with three-body forces:
  Binding energy and single-particle properties},\ }\href
  {https://doi.org/10.1103/PhysRevC.78.054003} {\bibfield  {journal} {\bibinfo
  {journal} {Phys. Rev. C}\ }\textbf {\bibinfo {volume} {78}},\ \bibinfo
  {pages} {054003} (\bibinfo {year} {2008})}\BibitemShut {NoStop}%
\bibitem [{\citenamefont {Barrett}\ \emph {et~al.}(2013)\citenamefont
  {Barrett}, \citenamefont {Navr{\'a}til},\ and\ \citenamefont
  {Vary}}]{barrett2013}%
  \BibitemOpen
  \bibfield  {author} {\bibinfo {author} {\bibfnamefont {B.~R.}\ \bibnamefont
  {Barrett}}, \bibinfo {author} {\bibfnamefont {P.}~\bibnamefont
  {Navr{\'a}til}},\ and\ \bibinfo {author} {\bibfnamefont {J.~P.}\ \bibnamefont
  {Vary}},\ }\bibfield  {title} {\bibinfo {title} {Ab initio no core shell
  model},\ }\href {https://doi.org/10.1016/j.ppnp.2012.10.003} {\bibfield
  {journal} {\bibinfo  {journal} {Prog. Part. Nucl. Phys.}\ }\textbf {\bibinfo
  {volume} {69}},\ \bibinfo {pages} {131 } (\bibinfo {year}
  {2013})}\BibitemShut {NoStop}%
\bibitem [{\citenamefont {Hagen}\ \emph {et~al.}(2014)\citenamefont {Hagen},
  \citenamefont {Papenbrock}, \citenamefont {Hjorth-Jensen},\ and\
  \citenamefont {Dean}}]{hagen2014}%
  \BibitemOpen
  \bibfield  {author} {\bibinfo {author} {\bibfnamefont {G.}~\bibnamefont
  {Hagen}}, \bibinfo {author} {\bibfnamefont {T.}~\bibnamefont {Papenbrock}},
  \bibinfo {author} {\bibfnamefont {M.}~\bibnamefont {Hjorth-Jensen}},\ and\
  \bibinfo {author} {\bibfnamefont {D.~J.}\ \bibnamefont {Dean}},\ }\bibfield
  {title} {\bibinfo {title} {Coupled-cluster computations of atomic nuclei},\
  }\href {https://doi.org/10.1088/0034-4885/77/9/096302} {\bibfield  {journal}
  {\bibinfo  {journal} {Rep. Prog. Phys.}\ }\textbf {\bibinfo {volume} {77}},\
  \bibinfo {pages} {096302} (\bibinfo {year} {2014})}\BibitemShut {NoStop}%
\bibitem [{\citenamefont {Hergert}\ \emph {et~al.}(2016)\citenamefont
  {Hergert}, \citenamefont {Bogner}, \citenamefont {Morris}, \citenamefont
  {Schwenk},\ and\ \citenamefont {Tsukiyama}}]{hergert2016}%
  \BibitemOpen
  \bibfield  {author} {\bibinfo {author} {\bibfnamefont {H.}~\bibnamefont
  {Hergert}}, \bibinfo {author} {\bibfnamefont {S.~K.}\ \bibnamefont {Bogner}},
  \bibinfo {author} {\bibfnamefont {T.~D.}\ \bibnamefont {Morris}}, \bibinfo
  {author} {\bibfnamefont {A.}~\bibnamefont {Schwenk}},\ and\ \bibinfo {author}
  {\bibfnamefont {K.}~\bibnamefont {Tsukiyama}},\ }\bibfield  {title} {\bibinfo
  {title} {The in-medium similarity renormalization group: A novel ab initio
  method for nuclei},\ }\href {https://doi.org/10.1016/j.physrep.2015.12.007}
  {\bibfield  {journal} {\bibinfo  {journal} {Phys. Rep.}\ }\textbf {\bibinfo
  {volume} {621}},\ \bibinfo {pages} {165 } (\bibinfo {year}
  {2016})}\BibitemShut {NoStop}%
\bibitem [{\citenamefont {Stroberg}\ \emph {et~al.}(2019)\citenamefont
  {Stroberg}, \citenamefont {Hergert}, \citenamefont {Bogner},\ and\
  \citenamefont {Holt}}]{stroberg2019}%
  \BibitemOpen
  \bibfield  {author} {\bibinfo {author} {\bibfnamefont {S.~R.}\ \bibnamefont
  {Stroberg}}, \bibinfo {author} {\bibfnamefont {H.}~\bibnamefont {Hergert}},
  \bibinfo {author} {\bibfnamefont {S.~K.}\ \bibnamefont {Bogner}},\ and\
  \bibinfo {author} {\bibfnamefont {J.~D.}\ \bibnamefont {Holt}},\ }\bibfield
  {title} {\bibinfo {title} {{Nonempirical Interactions for the Nuclear Shell
  Model: An Update}},\ }\href
  {https://doi.org/10.1146/annurev-nucl-101917-021120} {\bibfield  {journal}
  {\bibinfo  {journal} {Annu. Rev. Nucl. Part. Sci.}\ }\textbf {\bibinfo
  {volume} {69}},\ \bibinfo {pages} {307} (\bibinfo {year} {2019})}\BibitemShut
  {NoStop}%
\bibitem [{\citenamefont {Hebeler}\ \emph {et~al.}(2015)\citenamefont
  {Hebeler}, \citenamefont {Krebs}, \citenamefont {Epelbaum}, \citenamefont
  {Golak},\ and\ \citenamefont {Skibi\ifmmode~\acute{n}\else
  \'{n}\fi{}ski}}]{hebeler2015b}%
  \BibitemOpen
  \bibfield  {author} {\bibinfo {author} {\bibfnamefont {K.}~\bibnamefont
  {Hebeler}}, \bibinfo {author} {\bibfnamefont {H.}~\bibnamefont {Krebs}},
  \bibinfo {author} {\bibfnamefont {E.}~\bibnamefont {Epelbaum}}, \bibinfo
  {author} {\bibfnamefont {J.}~\bibnamefont {Golak}},\ and\ \bibinfo {author}
  {\bibfnamefont {R.}~\bibnamefont {Skibi\ifmmode~\acute{n}\else
  \'{n}\fi{}ski}},\ }\bibfield  {title} {\bibinfo {title} {Efficient
  calculation of chiral three-nucleon forces up to
  {${\mathrm{N}}^{3}\text{LO}$} for \textit{ab initio} studies},\ }\href
  {https://doi.org/10.1103/PhysRevC.91.044001} {\bibfield  {journal} {\bibinfo
  {journal} {Phys. Rev. C}\ }\textbf {\bibinfo {volume} {91}},\ \bibinfo
  {pages} {044001} (\bibinfo {year} {2015})}\BibitemShut {NoStop}%
\bibitem [{\citenamefont {Hebeler}(2021)}]{hebeler2021}%
  \BibitemOpen
  \bibfield  {author} {\bibinfo {author} {\bibfnamefont {K.}~\bibnamefont
  {Hebeler}},\ }\bibfield  {title} {\bibinfo {title} {Three-nucleon forces:
  Implementation and applications to atomic nuclei and dense matter},\ }\href
  {https://doi.org/10.1016/j.physrep.2020.08.009} {\bibfield  {journal}
  {\bibinfo  {journal} {Phys. Rep.}\ }\textbf {\bibinfo {volume} {890}},\
  \bibinfo {pages} {1} (\bibinfo {year} {2021})}\BibitemShut {NoStop}%
\bibitem [{\citenamefont {Miyagi}\ \emph {et~al.}(2022)\citenamefont {Miyagi},
  \citenamefont {Stroberg}, \citenamefont {Navr\'atil}, \citenamefont
  {Hebeler},\ and\ \citenamefont {Holt}}]{takayuki2022}%
  \BibitemOpen
  \bibfield  {author} {\bibinfo {author} {\bibfnamefont {T.}~\bibnamefont
  {Miyagi}}, \bibinfo {author} {\bibfnamefont {S.~R.}\ \bibnamefont
  {Stroberg}}, \bibinfo {author} {\bibfnamefont {P.}~\bibnamefont
  {Navr\'atil}}, \bibinfo {author} {\bibfnamefont {K.}~\bibnamefont
  {Hebeler}},\ and\ \bibinfo {author} {\bibfnamefont {J.~D.}\ \bibnamefont
  {Holt}},\ }\bibfield  {title} {\bibinfo {title} {Converged ab initio
  calculations of heavy nuclei},\ }\href
  {https://doi.org/10.1103/PhysRevC.105.014302} {\bibfield  {journal} {\bibinfo
   {journal} {Phys. Rev. C}\ }\textbf {\bibinfo {volume} {105}},\ \bibinfo
  {pages} {014302} (\bibinfo {year} {2022})}\BibitemShut {NoStop}%
\bibitem [{\citenamefont {Tichai}\ \emph {et~al.}(2024)\citenamefont {Tichai},
  \citenamefont {Arthuis}, \citenamefont {Hebeler}, \citenamefont {Heinz},
  \citenamefont {Hoppe}, \citenamefont {Miyagi}, \citenamefont {Schwenk},\ and\
  \citenamefont {Zurek}}]{tichai2024}%
  \BibitemOpen
  \bibfield  {author} {\bibinfo {author} {\bibfnamefont {A.}~\bibnamefont
  {Tichai}}, \bibinfo {author} {\bibfnamefont {P.}~\bibnamefont {Arthuis}},
  \bibinfo {author} {\bibfnamefont {K.}~\bibnamefont {Hebeler}}, \bibinfo
  {author} {\bibfnamefont {M.}~\bibnamefont {Heinz}}, \bibinfo {author}
  {\bibfnamefont {J.}~\bibnamefont {Hoppe}}, \bibinfo {author} {\bibfnamefont
  {T.}~\bibnamefont {Miyagi}}, \bibinfo {author} {\bibfnamefont
  {A.}~\bibnamefont {Schwenk}},\ and\ \bibinfo {author} {\bibfnamefont
  {L.}~\bibnamefont {Zurek}},\ }\bibfield  {title} {\bibinfo {title}
  {Randomized low-rank decompositions of nuclear three-body interactions},\
  }\href {https://doi.org/10.1103/PhysRevResearch.6.043331} {\bibfield
  {journal} {\bibinfo  {journal} {Phys. Rev. Res.}\ }\textbf {\bibinfo {volume}
  {6}},\ \bibinfo {pages} {043331} (\bibinfo {year} {2024})}\BibitemShut
  {NoStop}%
\bibitem [{\citenamefont {Hagen}\ \emph
  {et~al.}(2007{\natexlab{a}})\citenamefont {Hagen}, \citenamefont
  {Papenbrock}, \citenamefont {Dean}, \citenamefont {Schwenk}, \citenamefont
  {Nogga}, \citenamefont {W\l{}och},\ and\ \citenamefont
  {Piecuch}}]{hagen2007a}%
  \BibitemOpen
  \bibfield  {author} {\bibinfo {author} {\bibfnamefont {G.}~\bibnamefont
  {Hagen}}, \bibinfo {author} {\bibfnamefont {T.}~\bibnamefont {Papenbrock}},
  \bibinfo {author} {\bibfnamefont {D.~J.}\ \bibnamefont {Dean}}, \bibinfo
  {author} {\bibfnamefont {A.}~\bibnamefont {Schwenk}}, \bibinfo {author}
  {\bibfnamefont {A.}~\bibnamefont {Nogga}}, \bibinfo {author} {\bibfnamefont
  {M.}~\bibnamefont {W\l{}och}},\ and\ \bibinfo {author} {\bibfnamefont
  {P.}~\bibnamefont {Piecuch}},\ }\bibfield  {title} {\bibinfo {title}
  {{Coupled-cluster theory for three-body Hamiltonians}},\ }\href
  {https://doi.org/10.1103/PhysRevC.76.034302} {\bibfield  {journal} {\bibinfo
  {journal} {Phys. Rev. C}\ }\textbf {\bibinfo {volume} {76}},\ \bibinfo
  {pages} {034302} (\bibinfo {year} {2007}{\natexlab{a}})}\BibitemShut
  {NoStop}%
\bibitem [{\citenamefont {Roth}\ \emph {et~al.}(2012)\citenamefont {Roth},
  \citenamefont {Binder}, \citenamefont {Vobig}, \citenamefont {Calci},
  \citenamefont {Langhammer},\ and\ \citenamefont {Navr\'atil}}]{roth2012}%
  \BibitemOpen
  \bibfield  {author} {\bibinfo {author} {\bibfnamefont {R.}~\bibnamefont
  {Roth}}, \bibinfo {author} {\bibfnamefont {S.}~\bibnamefont {Binder}},
  \bibinfo {author} {\bibfnamefont {K.}~\bibnamefont {Vobig}}, \bibinfo
  {author} {\bibfnamefont {A.}~\bibnamefont {Calci}}, \bibinfo {author}
  {\bibfnamefont {J.}~\bibnamefont {Langhammer}},\ and\ \bibinfo {author}
  {\bibfnamefont {P.}~\bibnamefont {Navr\'atil}},\ }\bibfield  {title}
  {\bibinfo {title} {{Medium-Mass Nuclei with Normal-Ordered Chiral
  $NN\mathbf{+}3N$ Interactions}},\ }\href
  {https://doi.org/10.1103/PhysRevLett.109.052501} {\bibfield  {journal}
  {\bibinfo  {journal} {Phys. Rev. Lett.}\ }\textbf {\bibinfo {volume} {109}},\
  \bibinfo {pages} {052501} (\bibinfo {year} {2012})}\BibitemShut {NoStop}%
\bibitem [{\citenamefont {Ripoche}\ \emph {et~al.}(2020)\citenamefont
  {Ripoche}, \citenamefont {Tichai},\ and\ \citenamefont
  {Duguet}}]{ripoche2020}%
  \BibitemOpen
  \bibfield  {author} {\bibinfo {author} {\bibfnamefont {J.}~\bibnamefont
  {Ripoche}}, \bibinfo {author} {\bibfnamefont {A.}~\bibnamefont {Tichai}},\
  and\ \bibinfo {author} {\bibfnamefont {T.}~\bibnamefont {Duguet}},\
  }\bibfield  {title} {\bibinfo {title} {{Normal-ordered $k$-body approximation
  in particle-number-breaking theories}},\ }\href
  {https://doi.org/10.1140/epja/s10050-020-00045-8} {\bibfield  {journal}
  {\bibinfo  {journal} {Eur. Phys. J. A}\ }\textbf {\bibinfo {volume} {56}},\
  \bibinfo {pages} {40} (\bibinfo {year} {2020})}\BibitemShut {NoStop}%
\bibitem [{\citenamefont {Mayer}\ and\ \citenamefont
  {Jensen}(1955)}]{mayer1955}%
  \BibitemOpen
  \bibfield  {author} {\bibinfo {author} {\bibfnamefont {M.~G.}\ \bibnamefont
  {Mayer}}\ and\ \bibinfo {author} {\bibfnamefont {J.~H.~D.}\ \bibnamefont
  {Jensen}},\ }\href@noop {} {\emph {\bibinfo {title} {Elementary Theory of
  Nuclear Shell Structure}}}\ (\bibinfo  {publisher} {Wiley \& Sons},\ \bibinfo
  {address} {New York},\ \bibinfo {year} {1955})\BibitemShut {NoStop}%
\bibitem [{\citenamefont {Binder}\ \emph {et~al.}(2014)\citenamefont {Binder},
  \citenamefont {Langhammer}, \citenamefont {Calci},\ and\ \citenamefont
  {Roth}}]{binder2013b}%
  \BibitemOpen
  \bibfield  {author} {\bibinfo {author} {\bibfnamefont {S.}~\bibnamefont
  {Binder}}, \bibinfo {author} {\bibfnamefont {J.}~\bibnamefont {Langhammer}},
  \bibinfo {author} {\bibfnamefont {A.}~\bibnamefont {Calci}},\ and\ \bibinfo
  {author} {\bibfnamefont {R.}~\bibnamefont {Roth}},\ }\bibfield  {title}
  {\bibinfo {title} {Ab initio path to heavy nuclei},\ }\href
  {https://doi.org/10.1016/j.physletb.2014.07.010} {\bibfield  {journal}
  {\bibinfo  {journal} {Phys. Lett. B}\ }\textbf {\bibinfo {volume} {736}},\
  \bibinfo {pages} {119 } (\bibinfo {year} {2014})}\BibitemShut {NoStop}%
\bibitem [{\citenamefont {Cipollone}\ \emph {et~al.}(2013)\citenamefont
  {Cipollone}, \citenamefont {Barbieri},\ and\ \citenamefont
  {Navr\'atil}}]{cipollone2013}%
  \BibitemOpen
  \bibfield  {author} {\bibinfo {author} {\bibfnamefont {A.}~\bibnamefont
  {Cipollone}}, \bibinfo {author} {\bibfnamefont {C.}~\bibnamefont
  {Barbieri}},\ and\ \bibinfo {author} {\bibfnamefont {P.}~\bibnamefont
  {Navr\'atil}},\ }\bibfield  {title} {\bibinfo {title} {Isotopic chains around
  oxygen from evolved chiral two- and three-nucleon interactions},\ }\href
  {https://doi.org/10.1103/PhysRevLett.111.062501} {\bibfield  {journal}
  {\bibinfo  {journal} {Phys. Rev. Lett.}\ }\textbf {\bibinfo {volume} {111}},\
  \bibinfo {pages} {062501} (\bibinfo {year} {2013})}\BibitemShut {NoStop}%
\bibitem [{\citenamefont {Carbone}\ \emph
  {et~al.}(2013{\natexlab{a}})\citenamefont {Carbone}, \citenamefont
  {Cipollone}, \citenamefont {Barbieri}, \citenamefont {Rios},\ and\
  \citenamefont {Polls}}]{carbone2013b}%
  \BibitemOpen
  \bibfield  {author} {\bibinfo {author} {\bibfnamefont {A.}~\bibnamefont
  {Carbone}}, \bibinfo {author} {\bibfnamefont {A.}~\bibnamefont {Cipollone}},
  \bibinfo {author} {\bibfnamefont {C.}~\bibnamefont {Barbieri}}, \bibinfo
  {author} {\bibfnamefont {A.}~\bibnamefont {Rios}},\ and\ \bibinfo {author}
  {\bibfnamefont {A.}~\bibnamefont {Polls}},\ }\bibfield  {title} {\bibinfo
  {title} {Self-consistent {G}reen's functions formalism with three-body
  interactions},\ }\href {https://doi.org/10.1103/PhysRevC.88.054326}
  {\bibfield  {journal} {\bibinfo  {journal} {Phys. Rev. C}\ }\textbf {\bibinfo
  {volume} {88}},\ \bibinfo {pages} {054326} (\bibinfo {year}
  {2013}{\natexlab{a}})}\BibitemShut {NoStop}%
\bibitem [{\citenamefont {Heinz}\ \emph {et~al.}(2021)\citenamefont {Heinz},
  \citenamefont {Tichai}, \citenamefont {Hoppe}, \citenamefont {Hebeler},\ and\
  \citenamefont {Schwenk}}]{heinz2021}%
  \BibitemOpen
  \bibfield  {author} {\bibinfo {author} {\bibfnamefont {M.}~\bibnamefont
  {Heinz}}, \bibinfo {author} {\bibfnamefont {A.}~\bibnamefont {Tichai}},
  \bibinfo {author} {\bibfnamefont {J.}~\bibnamefont {Hoppe}}, \bibinfo
  {author} {\bibfnamefont {K.}~\bibnamefont {Hebeler}},\ and\ \bibinfo {author}
  {\bibfnamefont {A.}~\bibnamefont {Schwenk}},\ }\bibfield  {title} {\bibinfo
  {title} {In-medium similarity renormalization group with three-body
  operators},\ }\href {https://doi.org/10.1103/PhysRevC.103.044318} {\bibfield
  {journal} {\bibinfo  {journal} {Phys. Rev. C}\ }\textbf {\bibinfo {volume}
  {103}},\ \bibinfo {pages} {044318} (\bibinfo {year} {2021})}\BibitemShut
  {NoStop}%
\bibitem [{\citenamefont {Arthuis}\ \emph {et~al.}(2023)\citenamefont
  {Arthuis}, \citenamefont {Barbieri}, \citenamefont {Pederiva},\ and\
  \citenamefont {Roggero}}]{arthuis2023}%
  \BibitemOpen
  \bibfield  {author} {\bibinfo {author} {\bibfnamefont {P.}~\bibnamefont
  {Arthuis}}, \bibinfo {author} {\bibfnamefont {C.}~\bibnamefont {Barbieri}},
  \bibinfo {author} {\bibfnamefont {F.}~\bibnamefont {Pederiva}},\ and\
  \bibinfo {author} {\bibfnamefont {A.}~\bibnamefont {Roggero}},\ }\bibfield
  {title} {\bibinfo {title} {Quantum {M}onte {C}arlo calculations in
  configuration space with three-nucleon forces},\ }\href
  {https://doi.org/10.1103/PhysRevC.107.044303} {\bibfield  {journal} {\bibinfo
   {journal} {Phys. Rev. C}\ }\textbf {\bibinfo {volume} {107}},\ \bibinfo
  {pages} {044303} (\bibinfo {year} {2023})}\BibitemShut {NoStop}%
\bibitem [{\citenamefont {Holt}\ \emph {et~al.}(2009)\citenamefont {Holt},
  \citenamefont {Kaiser},\ and\ \citenamefont {Weise}}]{holt2009}%
  \BibitemOpen
  \bibfield  {author} {\bibinfo {author} {\bibfnamefont {J.~W.}\ \bibnamefont
  {Holt}}, \bibinfo {author} {\bibfnamefont {N.}~\bibnamefont {Kaiser}},\ and\
  \bibinfo {author} {\bibfnamefont {W.}~\bibnamefont {Weise}},\ }\bibfield
  {title} {\bibinfo {title} {Chiral three-nucleon interaction and the
  $^{14}\mathrm{C}$-dating $\ensuremath{\beta}$ decay},\ }\href
  {https://doi.org/10.1103/PhysRevC.79.054331} {\bibfield  {journal} {\bibinfo
  {journal} {Phys. Rev. C}\ }\textbf {\bibinfo {volume} {79}},\ \bibinfo
  {pages} {054331} (\bibinfo {year} {2009})}\BibitemShut {NoStop}%
\bibitem [{\citenamefont {Hagen}\ \emph
  {et~al.}(2012{\natexlab{a}})\citenamefont {Hagen}, \citenamefont
  {Hjorth-Jensen}, \citenamefont {Jansen}, \citenamefont {Machleidt},\ and\
  \citenamefont {Papenbrock}}]{hagen2012a}%
  \BibitemOpen
  \bibfield  {author} {\bibinfo {author} {\bibfnamefont {G.}~\bibnamefont
  {Hagen}}, \bibinfo {author} {\bibfnamefont {M.}~\bibnamefont
  {Hjorth-Jensen}}, \bibinfo {author} {\bibfnamefont {G.~R.}\ \bibnamefont
  {Jansen}}, \bibinfo {author} {\bibfnamefont {R.}~\bibnamefont {Machleidt}},\
  and\ \bibinfo {author} {\bibfnamefont {T.}~\bibnamefont {Papenbrock}},\
  }\bibfield  {title} {\bibinfo {title} {Continuum effects and three-nucleon
  forces in neutron-rich oxygen isotopes},\ }\href
  {https://doi.org/10.1103/PhysRevLett.108.242501} {\bibfield  {journal}
  {\bibinfo  {journal} {Phys. Rev. Lett.}\ }\textbf {\bibinfo {volume} {108}},\
  \bibinfo {pages} {242501} (\bibinfo {year} {2012}{\natexlab{a}})}\BibitemShut
  {NoStop}%
\bibitem [{\citenamefont {Hagen}\ \emph
  {et~al.}(2012{\natexlab{b}})\citenamefont {Hagen}, \citenamefont
  {Hjorth-Jensen}, \citenamefont {Jansen}, \citenamefont {Machleidt},\ and\
  \citenamefont {Papenbrock}}]{hagen2012b}%
  \BibitemOpen
  \bibfield  {author} {\bibinfo {author} {\bibfnamefont {G.}~\bibnamefont
  {Hagen}}, \bibinfo {author} {\bibfnamefont {M.}~\bibnamefont
  {Hjorth-Jensen}}, \bibinfo {author} {\bibfnamefont {G.~R.}\ \bibnamefont
  {Jansen}}, \bibinfo {author} {\bibfnamefont {R.}~\bibnamefont {Machleidt}},\
  and\ \bibinfo {author} {\bibfnamefont {T.}~\bibnamefont {Papenbrock}},\
  }\bibfield  {title} {\bibinfo {title} {Evolution of shell structure in
  neutron-rich calcium isotopes},\ }\href
  {https://doi.org/10.1103/PhysRevLett.109.032502} {\bibfield  {journal}
  {\bibinfo  {journal} {Phys. Rev. Lett.}\ }\textbf {\bibinfo {volume} {109}},\
  \bibinfo {pages} {032502} (\bibinfo {year} {2012}{\natexlab{b}})}\BibitemShut
  {NoStop}%
\bibitem [{\citenamefont {Carbone}\ \emph
  {et~al.}(2013{\natexlab{b}})\citenamefont {Carbone}, \citenamefont {Polls},\
  and\ \citenamefont {Rios}}]{carbone2013}%
  \BibitemOpen
  \bibfield  {author} {\bibinfo {author} {\bibfnamefont {A.}~\bibnamefont
  {Carbone}}, \bibinfo {author} {\bibfnamefont {A.}~\bibnamefont {Polls}},\
  and\ \bibinfo {author} {\bibfnamefont {A.}~\bibnamefont {Rios}},\ }\bibfield
  {title} {\bibinfo {title} {Symmetric nuclear matter with chiral three-nucleon
  forces in the self-consistent {G}reen's functions approach},\ }\href
  {https://doi.org/10.1103/PhysRevC.88.044302} {\bibfield  {journal} {\bibinfo
  {journal} {Phys. Rev. C}\ }\textbf {\bibinfo {volume} {88}},\ \bibinfo
  {pages} {044302} (\bibinfo {year} {2013}{\natexlab{b}})}\BibitemShut
  {NoStop}%
\bibitem [{\citenamefont {Hammer}\ \emph {et~al.}(2020)\citenamefont {Hammer},
  \citenamefont {K\"onig},\ and\ \citenamefont {van Kolck}}]{Hammer:2019poc}%
  \BibitemOpen
  \bibfield  {author} {\bibinfo {author} {\bibfnamefont {H.~W.}\ \bibnamefont
  {Hammer}}, \bibinfo {author} {\bibfnamefont {S.}~\bibnamefont {K\"onig}},\
  and\ \bibinfo {author} {\bibfnamefont {U.}~\bibnamefont {van Kolck}},\
  }\bibfield  {title} {\bibinfo {title} {{Nuclear effective field theory:
  Status and perspectives}},\ }\href
  {https://doi.org/10.1103/RevModPhys.92.025004} {\bibfield  {journal}
  {\bibinfo  {journal} {Rev. Mod. Phys.}\ }\textbf {\bibinfo {volume} {92}},\
  \bibinfo {pages} {025004} (\bibinfo {year} {2020})}\BibitemShut {NoStop}%
\bibitem [{\citenamefont {Lee}(2009)}]{lee2009}%
  \BibitemOpen
  \bibfield  {author} {\bibinfo {author} {\bibfnamefont {D.}~\bibnamefont
  {Lee}},\ }\bibfield  {title} {\bibinfo {title} {Lattice simulations for few-
  and many-body systems},\ }\href {https://doi.org/10.1016/j.ppnp.2008.12.001}
  {\bibfield  {journal} {\bibinfo  {journal} {Prog. Part. Nucl. Phys.}\
  }\textbf {\bibinfo {volume} {63}},\ \bibinfo {pages} {117 } (\bibinfo {year}
  {2009})}\BibitemShut {NoStop}%
\bibitem [{\citenamefont {L{\"a}hde}\ and\ \citenamefont
  {Mei{\ss}ner}(2019)}]{lahde2019}%
  \BibitemOpen
  \bibfield  {author} {\bibinfo {author} {\bibfnamefont {T.~A.}\ \bibnamefont
  {L{\"a}hde}}\ and\ \bibinfo {author} {\bibfnamefont {U.-G.}\ \bibnamefont
  {Mei{\ss}ner}},\ }\href {https://doi.org/10.1007/978-3-030-14189-9} {\emph
  {\bibinfo {title} {Nuclear Lattice Effective Field Theory}}},\ Lecture Notes
  in Physics\ (\bibinfo  {publisher} {Springer},\ \bibinfo {address} {Cham,
  Switzerland},\ \bibinfo {year} {2019})\BibitemShut {NoStop}%
\bibitem [{\citenamefont {Epelbaum}\ \emph {et~al.}(2012)\citenamefont
  {Epelbaum}, \citenamefont {Krebs}, \citenamefont {L\"ahde}, \citenamefont
  {Lee},\ and\ \citenamefont {Mei\ss{}ner}}]{epelbaum2012}%
  \BibitemOpen
  \bibfield  {author} {\bibinfo {author} {\bibfnamefont {E.}~\bibnamefont
  {Epelbaum}}, \bibinfo {author} {\bibfnamefont {H.}~\bibnamefont {Krebs}},
  \bibinfo {author} {\bibfnamefont {T.~A.}\ \bibnamefont {L\"ahde}}, \bibinfo
  {author} {\bibfnamefont {D.}~\bibnamefont {Lee}},\ and\ \bibinfo {author}
  {\bibfnamefont {U.-G.}\ \bibnamefont {Mei\ss{}ner}},\ }\bibfield  {title}
  {\bibinfo {title} {Structure and rotations of the {H}oyle state},\ }\href
  {https://doi.org/10.1103/PhysRevLett.109.252501} {\bibfield  {journal}
  {\bibinfo  {journal} {Phys. Rev. Lett.}\ }\textbf {\bibinfo {volume} {109}},\
  \bibinfo {pages} {252501} (\bibinfo {year} {2012})}\BibitemShut {NoStop}%
\bibitem [{\citenamefont {Epelbaum}\ \emph {et~al.}(2014)\citenamefont
  {Epelbaum}, \citenamefont {Krebs}, \citenamefont {L\"ahde}, \citenamefont
  {Lee}, \citenamefont {Mei\ss{}ner},\ and\ \citenamefont
  {Rupak}}]{epelbaum2014}%
  \BibitemOpen
  \bibfield  {author} {\bibinfo {author} {\bibfnamefont {E.}~\bibnamefont
  {Epelbaum}}, \bibinfo {author} {\bibfnamefont {H.}~\bibnamefont {Krebs}},
  \bibinfo {author} {\bibfnamefont {T.~A.}\ \bibnamefont {L\"ahde}}, \bibinfo
  {author} {\bibfnamefont {D.}~\bibnamefont {Lee}}, \bibinfo {author}
  {\bibfnamefont {U.-G.}\ \bibnamefont {Mei\ss{}ner}},\ and\ \bibinfo {author}
  {\bibfnamefont {G.}~\bibnamefont {Rupak}},\ }\bibfield  {title} {\bibinfo
  {title} {Ab initio calculation of the spectrum and structure of {$^{16}$O}},\
  }\href {https://doi.org/10.1103/PhysRevLett.112.102501} {\bibfield  {journal}
  {\bibinfo  {journal} {Phys. Rev. Lett.}\ }\textbf {\bibinfo {volume} {112}},\
  \bibinfo {pages} {102501} (\bibinfo {year} {2014})}\BibitemShut {NoStop}%
\bibitem [{\citenamefont {Roggero}\ \emph {et~al.}(2020)\citenamefont
  {Roggero}, \citenamefont {Li}, \citenamefont {Carlson}, \citenamefont
  {Gupta},\ and\ \citenamefont {Perdue}}]{roggero2020}%
  \BibitemOpen
  \bibfield  {author} {\bibinfo {author} {\bibfnamefont {A.}~\bibnamefont
  {Roggero}}, \bibinfo {author} {\bibfnamefont {A.~C.~Y.}\ \bibnamefont {Li}},
  \bibinfo {author} {\bibfnamefont {J.}~\bibnamefont {Carlson}}, \bibinfo
  {author} {\bibfnamefont {R.}~\bibnamefont {Gupta}},\ and\ \bibinfo {author}
  {\bibfnamefont {G.~N.}\ \bibnamefont {Perdue}},\ }\bibfield  {title}
  {\bibinfo {title} {Quantum computing for neutrino-nucleus scattering},\
  }\href {https://doi.org/10.1103/PhysRevD.101.074038} {\bibfield  {journal}
  {\bibinfo  {journal} {Phys. Rev. D}\ }\textbf {\bibinfo {volume} {101}},\
  \bibinfo {pages} {074038} (\bibinfo {year} {2020})}\BibitemShut {NoStop}%
\bibitem [{\citenamefont {Baroni}\ \emph {et~al.}(2022)\citenamefont {Baroni},
  \citenamefont {Carlson}, \citenamefont {Gupta}, \citenamefont {Li},
  \citenamefont {Perdue},\ and\ \citenamefont {Roggero}}]{baroni2022}%
  \BibitemOpen
  \bibfield  {author} {\bibinfo {author} {\bibfnamefont {A.}~\bibnamefont
  {Baroni}}, \bibinfo {author} {\bibfnamefont {J.}~\bibnamefont {Carlson}},
  \bibinfo {author} {\bibfnamefont {R.}~\bibnamefont {Gupta}}, \bibinfo
  {author} {\bibfnamefont {A.~C.~Y.}\ \bibnamefont {Li}}, \bibinfo {author}
  {\bibfnamefont {G.~N.}\ \bibnamefont {Perdue}},\ and\ \bibinfo {author}
  {\bibfnamefont {A.}~\bibnamefont {Roggero}},\ }\bibfield  {title} {\bibinfo
  {title} {Nuclear two point correlation functions on a quantum computer},\
  }\href {https://doi.org/10.1103/PhysRevD.105.074503} {\bibfield  {journal}
  {\bibinfo  {journal} {Phys. Rev. D}\ }\textbf {\bibinfo {volume} {105}},\
  \bibinfo {pages} {074503} (\bibinfo {year} {2022})}\BibitemShut {NoStop}%
\bibitem [{\citenamefont {Gu}\ \emph {et~al.}()\citenamefont {Gu},
  \citenamefont {Heinz}, \citenamefont {Kiss},\ and\ \citenamefont
  {Papenbrock}}]{gu2025}%
  \BibitemOpen
  \bibfield  {author} {\bibinfo {author} {\bibfnamefont {C.}~\bibnamefont
  {Gu}}, \bibinfo {author} {\bibfnamefont {M.}~\bibnamefont {Heinz}}, \bibinfo
  {author} {\bibfnamefont {O.}~\bibnamefont {Kiss}},\ and\ \bibinfo {author}
  {\bibfnamefont {T.}~\bibnamefont {Papenbrock}},\ }\href@noop {} {\bibinfo
  {title} {{Towards scalable quantum computations of atomic nuclei}}},\ \Eprint
  {https://arxiv.org/abs/2507.14690} {arXiv:2507.14690 [nucl-th]} \BibitemShut
  {NoStop}%
\bibitem [{\citenamefont {L{\"u}scher}(1986)}]{luscher1985}%
  \BibitemOpen
  \bibfield  {author} {\bibinfo {author} {\bibfnamefont {M.}~\bibnamefont
  {L{\"u}scher}},\ }\bibfield  {title} {\bibinfo {title} {{Volume Dependence of
  the Energy Spectrum in Massive Quantum Field Theories. 1. Stable Particle
  States}},\ }\href {https://doi.org/10.1007/BF01211589} {\bibfield  {journal}
  {\bibinfo  {journal} {Commun. Math. Phys.}\ }\textbf {\bibinfo {volume}
  {104}},\ \bibinfo {pages} {177} (\bibinfo {year} {1986})}\BibitemShut
  {NoStop}%
\bibitem [{\citenamefont {K{\"o}nig}\ and\ \citenamefont
  {Lee}(2018)}]{konig2017}%
  \BibitemOpen
  \bibfield  {author} {\bibinfo {author} {\bibfnamefont {S.}~\bibnamefont
  {K{\"o}nig}}\ and\ \bibinfo {author} {\bibfnamefont {D.}~\bibnamefont
  {Lee}},\ }\bibfield  {title} {\bibinfo {title} {Volume dependence of n-body
  bound states},\ }\href {https://doi.org/10.1016/j.physletb.2018.01.060}
  {\bibfield  {journal} {\bibinfo  {journal} {Phys. Lett. B}\ }\textbf
  {\bibinfo {volume} {779}},\ \bibinfo {pages} {9 } (\bibinfo {year}
  {2018})}\BibitemShut {NoStop}%
\bibitem [{\citenamefont {Bartlett}\ and\ \citenamefont
  {Musia\l{}}(2007)}]{bartlett2007}%
  \BibitemOpen
  \bibfield  {author} {\bibinfo {author} {\bibfnamefont {R.~J.}\ \bibnamefont
  {Bartlett}}\ and\ \bibinfo {author} {\bibfnamefont {M.}~\bibnamefont
  {Musia\l{}}},\ }\bibfield  {title} {\bibinfo {title} {Coupled-cluster theory
  in quantum chemistry},\ }\href {https://doi.org/10.1103/RevModPhys.79.291}
  {\bibfield  {journal} {\bibinfo  {journal} {Rev. Mod. Phys.}\ }\textbf
  {\bibinfo {volume} {79}},\ \bibinfo {pages} {291} (\bibinfo {year}
  {2007})}\BibitemShut {NoStop}%
\bibitem [{\citenamefont {Hagen}\ \emph {et~al.}(2009)\citenamefont {Hagen},
  \citenamefont {Papenbrock}, \citenamefont {Dean}, \citenamefont
  {Hjorth-Jensen},\ and\ \citenamefont {Asokan}}]{hagen2009b}%
  \BibitemOpen
  \bibfield  {author} {\bibinfo {author} {\bibfnamefont {G.}~\bibnamefont
  {Hagen}}, \bibinfo {author} {\bibfnamefont {T.}~\bibnamefont {Papenbrock}},
  \bibinfo {author} {\bibfnamefont {D.~J.}\ \bibnamefont {Dean}}, \bibinfo
  {author} {\bibfnamefont {M.}~\bibnamefont {Hjorth-Jensen}},\ and\ \bibinfo
  {author} {\bibfnamefont {B.~V.}\ \bibnamefont {Asokan}},\ }\bibfield  {title}
  {\bibinfo {title} {\textit{Ab initio} computation of neutron-rich oxygen
  isotopes},\ }\href {https://doi.org/10.1103/PhysRevC.80.021306} {\bibfield
  {journal} {\bibinfo  {journal} {Phys. Rev. C}\ }\textbf {\bibinfo {volume}
  {80}},\ \bibinfo {pages} {021306} (\bibinfo {year} {2009})}\BibitemShut
  {NoStop}%
\bibitem [{\citenamefont {Sun}\ \emph {et~al.}(2022)\citenamefont {Sun},
  \citenamefont {Bell}, \citenamefont {Hagen},\ and\ \citenamefont
  {Papenbrock}}]{sun2022}%
  \BibitemOpen
  \bibfield  {author} {\bibinfo {author} {\bibfnamefont {Z.~H.}\ \bibnamefont
  {Sun}}, \bibinfo {author} {\bibfnamefont {C.~A.}\ \bibnamefont {Bell}},
  \bibinfo {author} {\bibfnamefont {G.}~\bibnamefont {Hagen}},\ and\ \bibinfo
  {author} {\bibfnamefont {T.}~\bibnamefont {Papenbrock}},\ }\bibfield  {title}
  {\bibinfo {title} {How to renormalize coupled cluster theory},\ }\href
  {https://doi.org/10.1103/PhysRevC.106.L061302} {\bibfield  {journal}
  {\bibinfo  {journal} {Phys. Rev. C}\ }\textbf {\bibinfo {volume} {106}},\
  \bibinfo {pages} {L061302} (\bibinfo {year} {2022})}\BibitemShut {NoStop}%
\bibitem [{\citenamefont {Sun}\ \emph {et~al.}(2025)\citenamefont {Sun},
  \citenamefont {Ekstr\"om}, \citenamefont {Forss\'en}, \citenamefont {Hagen},
  \citenamefont {Jansen},\ and\ \citenamefont {Papenbrock}}]{sun2025}%
  \BibitemOpen
  \bibfield  {author} {\bibinfo {author} {\bibfnamefont {Z.~H.}\ \bibnamefont
  {Sun}}, \bibinfo {author} {\bibfnamefont {A.}~\bibnamefont {Ekstr\"om}},
  \bibinfo {author} {\bibfnamefont {C.}~\bibnamefont {Forss\'en}}, \bibinfo
  {author} {\bibfnamefont {G.}~\bibnamefont {Hagen}}, \bibinfo {author}
  {\bibfnamefont {G.~R.}\ \bibnamefont {Jansen}},\ and\ \bibinfo {author}
  {\bibfnamefont {T.}~\bibnamefont {Papenbrock}},\ }\bibfield  {title}
  {\bibinfo {title} {Multiscale physics of atomic nuclei from first
  principles},\ }\href {https://doi.org/10.1103/PhysRevX.15.011028} {\bibfield
  {journal} {\bibinfo  {journal} {Phys. Rev. X}\ }\textbf {\bibinfo {volume}
  {15}},\ \bibinfo {pages} {011028} (\bibinfo {year} {2025})}\BibitemShut
  {NoStop}%
\bibitem [{\citenamefont {Hagen}\ \emph
  {et~al.}(2007{\natexlab{b}})\citenamefont {Hagen}, \citenamefont {Dean},
  \citenamefont {Hjorth-Jensen}, \citenamefont {Papenbrock},\ and\
  \citenamefont {Schwenk}}]{hagen2007b}%
  \BibitemOpen
  \bibfield  {author} {\bibinfo {author} {\bibfnamefont {G.}~\bibnamefont
  {Hagen}}, \bibinfo {author} {\bibfnamefont {D.~J.}\ \bibnamefont {Dean}},
  \bibinfo {author} {\bibfnamefont {M.}~\bibnamefont {Hjorth-Jensen}}, \bibinfo
  {author} {\bibfnamefont {T.}~\bibnamefont {Papenbrock}},\ and\ \bibinfo
  {author} {\bibfnamefont {A.}~\bibnamefont {Schwenk}},\ }\bibfield  {title}
  {\bibinfo {title} {Benchmark calculations for $^{3}\mathrm{H}$,
  $^{4}\mathrm{He}$, $^{16}\mathrm{O}$, and $^{40}\mathrm{Ca}$ with \textit{ab
  initio} coupled-cluster theory},\ }\href
  {https://doi.org/10.1103/PhysRevC.76.044305} {\bibfield  {journal} {\bibinfo
  {journal} {Phys. Rev. C}\ }\textbf {\bibinfo {volume} {76}},\ \bibinfo
  {pages} {044305} (\bibinfo {year} {2007}{\natexlab{b}})}\BibitemShut
  {NoStop}%
\bibitem [{\citenamefont {Bazak}\ \emph {et~al.}(2019)\citenamefont {Bazak},
  \citenamefont {Kirscher}, \citenamefont {K\"onig}, \citenamefont
  {Valderrama}, \citenamefont {Barnea},\ and\ \citenamefont {van
  Kolck}}]{bazak2019}%
  \BibitemOpen
  \bibfield  {author} {\bibinfo {author} {\bibfnamefont {B.}~\bibnamefont
  {Bazak}}, \bibinfo {author} {\bibfnamefont {J.}~\bibnamefont {Kirscher}},
  \bibinfo {author} {\bibfnamefont {S.}~\bibnamefont {K\"onig}}, \bibinfo
  {author} {\bibfnamefont {M.~P.}\ \bibnamefont {Valderrama}}, \bibinfo
  {author} {\bibfnamefont {N.}~\bibnamefont {Barnea}},\ and\ \bibinfo {author}
  {\bibfnamefont {U.}~\bibnamefont {van Kolck}},\ }\bibfield  {title} {\bibinfo
  {title} {Four-body scale in universal few-boson systems},\ }\href
  {https://doi.org/10.1103/PhysRevLett.122.143001} {\bibfield  {journal}
  {\bibinfo  {journal} {Phys. Rev. Lett.}\ }\textbf {\bibinfo {volume} {122}},\
  \bibinfo {pages} {143001} (\bibinfo {year} {2019})}\BibitemShut {NoStop}%
\bibitem [{\citenamefont {Rothman}\ \emph {et~al.}(2025)\citenamefont
  {Rothman}, \citenamefont {Johnson-Toth}, \citenamefont {Bonaiti},
  \citenamefont {Hagen}, \citenamefont {Heinz},\ and\ \citenamefont
  {Papenbrock}}]{rothman_zenodo2025}%
  \BibitemOpen
  \bibfield  {author} {\bibinfo {author} {\bibfnamefont {M.}~\bibnamefont
  {Rothman}}, \bibinfo {author} {\bibfnamefont {B.}~\bibnamefont
  {Johnson-Toth}}, \bibinfo {author} {\bibfnamefont {F.}~\bibnamefont
  {Bonaiti}}, \bibinfo {author} {\bibfnamefont {G.}~\bibnamefont {Hagen}},
  \bibinfo {author} {\bibfnamefont {M.}~\bibnamefont {Heinz}},\ and\ \bibinfo
  {author} {\bibfnamefont {T.}~\bibnamefont {Papenbrock}},\ }\bibfield  {title}
  {\bibinfo {title} {Data: Exactness of the normal-ordered two-body truncation
  of three-nucleon forces},\ }\href {https://doi.org/10.5281/zenodo.16729363}
  {10.5281/zenodo.16729363} (\bibinfo {year} {2025})\BibitemShut {NoStop}%
\end{thebibliography}%
\end{document}